\documentclass[longauth]{aa}  
\usepackage{rotating}
\usepackage{amsmath}
\usepackage[colorlinks=true, citecolor=blue, urlcolor=blue]{hyperref}

\usepackage{graphicx}
\usepackage{txfonts}
\usepackage{amssymb}
\usepackage{lscape}
\usepackage[bf,nooneline]{subfigure}
\bibpunct{(}{)}{;}{a}{}{,}

\begin{document}

\title{VIMOS Public Extragalactic Redshift Survey (VIPERS)}
  \subtitle{The distinct build-up of dense and normal massive passive galaxies 
\thanks{based on observations collected at the European Southern Observatory, Cerro Paranal, Chile, using the Very Large Telescope under programs 182.A-0886 and partly 070.A-9007.
Also based on observations obtained with MegaPrime/MegaCam, a joint project of CFHT and CEA/DAPNIA, at the Canada-France-Hawaii Telescope (CFHT), which is operated by the
National Research Council (NRC) of Canada, the Institut National des Sciences de l’Univers of the Centre National de la Recherche Scientifique (CNRS) of France, and the University of Hawaii. This work is based in part on data products produced at TERAPIX and the Canadian Astronomy Data Centre as part of the Canada-France-Hawaii Telescope Legacy Survey, a collaborative project of NRC and CNRS. The VIPERS web site is http://www.vipers.inaf.it/. }}
\author{
A. Gargiulo\inst{\ref{iasf-mi}}
\and M.~Bolzonella\inst{\ref{oabo}} 
\and M.~Scodeggio\inst{\ref{iasf-mi}}
\and J.~Krywult\inst{\ref{kielce}}
\and G.~De Lucia\inst{\ref{oats}}
\and L.~Guzzo\inst{\ref{brera},\ref{unimi}}     
\and B.~Garilli\inst{\ref{iasf-mi}}          
\and B.~R.~Granett\inst{\ref{brera},\ref{unimi}}                                                      
\and S.~de la Torre\inst{\ref{lam}}       
\and U.~Abbas\inst{\ref{oa-to}}
\and C.~Adami\inst{\ref{lam}}
\and S.~Arnouts\inst{\ref{lam}} 
\and D.~Bottini\inst{\ref{iasf-mi}}
\and A.~Cappi\inst{\ref{oabo},\ref{nice}}
\and O.~Cucciati\inst{\ref{oabo},\ref{unibo}}           
\and I.~Davidzon\inst{\ref{lam},\ref{oabo}}   
\and P.~Franzetti\inst{\ref{iasf-mi}}   
\and A.~Fritz\inst{\ref{iasf-mi}}      
\and C.~Haines\inst{\ref{brera}}
\and A.~J.~Hawken\inst{\ref{brera},\ref{unimi}} 
\and A.~Iovino\inst{\ref{brera}}
\and V.~Le Brun\inst{\ref{lam}}
\and O.~Le F\`evre\inst{\ref{lam}}
\and D.~Maccagni\inst{\ref{iasf-mi}}
\and K.~Ma{\l}ek\inst{\ref{warsaw-nucl}}
\and F.~Marulli\inst{\ref{unibo},\ref{infn-bo},\ref{oabo}} 
\and T.~Moutard\inst{\ref{halifax},\ref{lam}}  
\and M.~Polletta\inst{\ref{iasf-mi},\ref{marseille-uni},\ref{toulouse}}
\and A.~Pollo\inst{\ref{warsaw-nucl},\ref{krakow}}
\and L.A.M.~Tasca\inst{\ref{lam}}
\and R.~Tojeiro\inst{\ref{st-andrews}}  
\and D.~Vergani\inst{\ref{iasf-bo}}
\and A.~Zanichelli\inst{\ref{ira-bo}}
\and G.~Zamorani\inst{\ref{oabo}}
\and J.~Bel\inst{\ref{cpt}}
\and E.~Branchini\inst{\ref{roma3},\ref{infn-roma3},\ref{oa-roma}}
\and J.~Coupon\inst{\ref{geneva}}
\and O.~Ilbert\inst{\ref{lam}}
\and L.~Moscardini\inst{\ref{unibo},\ref{infn-bo},\ref{oabo}}
}
   \offprints{Adriana Gargiulo \\ \email{adriana@lambrate.inaf.it}}
\institute{
INAF - Istituto di Astrofisica Spaziale e Fisica Cosmica Milano, via Bassini 15, 20133 Milano, Italy \label{iasf-mi}%3
\and INAF - Osservatorio Astronomico di Bologna, via Ranzani 1, I-40127, Bologna, Italy \label{oabo} %4
\and Institute of Physics, Jan Kochanowski University, ul. Swietokrzyska 15, 25-406 Kielce, Poland \label{kielce}%15
\and INAF - Osservatorio Astronomico di Trieste, via G. B. Tiepolo 11, 34143 Trieste, Italy \label{oats}%13
\and INAF - Osservatorio Astronomico di Brera, Via Brera 28, 20122 Milano --  via E. Bianchi 46, 23807 Merate, Italy \label{brera} %1
\and  Universit\`{a} degli Studi di Milano, via G. Celoria 16, 20133 Milano, Italy \label{unimi}%2
\and Aix Marseille Univ, CNRS, LAM, Laboratoire d'Astrophysique de Marseille, Marseille, France  \label{lam}%5
\and INAF - Osservatorio Astrofisico di Torino, 10025 Pino Torinese, Italy \label{oa-to}%5
\and Laboratoire Lagrange, UMR7293, Universit\'e de Nice Sophia Antipolis, CNRS, Observatoire de la C\^ote d'Azur, 06300 Nice, France \label{nice}%
\and Dipartimento di Fisica e Astronomia - Alma Mater Studiorum Universit\`{a} di Bologna, viale Berti Pichat 6/2, I-40127 Bologna, Italy \label{unibo}%17
\and National Centre for Nuclear Research, ul. Hoza 69, 00-681 Warszawa, Poland \label{warsaw-nucl}%23
\and INFN, Sezione di Bologna, viale Berti Pichat 6/2, I-40127 Bologna, Italy \label{infn-bo}%18
\and Department of Astronomy $\&$ Physics, Saint Mary's University, 923 Robie Street, Halifax, Nova Scotia, B3H 3C3, Canada \label{halifax}
\and Aix-Marseille UniversitÃ©, Jardin du Pharo, 58 bd Charles Livon, F-13284 Marseille cedex 7, France \label{marseille-uni}
\and IRAP,  9 av. du colonel Roche, BP 44346, F-31028 Toulouse cedex 4, France \label{toulouse} 
\and Astronomical Observatory of the Jagiellonian University, Orla 171, 30-001 Cracow, Poland \label{krakow} %22
\and School of Physics and Astronomy, University of St Andrews, St Andrews KY16 9SS, UK \label{st-andrews}%11
\and INAF - Istituto di Astrofisica Spaziale e Fisica Cosmica Bologna, via Gobetti 101, I-40129 Bologna, Italy \label{iasf-bo}%25
\and INAF - Istituto di Radioastronomia, via Gobetti 101, I-40129, Bologna, Italy \label{ira-bo}%26
\and Canada-France-Hawaii Telescope, 65--1238 Mamalahoa Highway, Kamuela, HI 96743, USA \label{cfht}%6
\and Aix Marseille Univ, Univ Toulon, CNRS, CPT, Marseille, France \label{cpt}%7
\and Dipartimento di Matematica e Fisica, Universit\`{a} degli Studi Roma Tre, via della Vasca Navale 84, 00146 Roma, Italy\label{roma3} %10
\and INFN, Sezione di Roma Tre, via della Vasca Navale 84, I-00146 Roma, Italy \label{infn-roma3}%28
\and INAF - Osservatorio Astronomico di Roma, via Frascati 33, I-00040 Monte Porzio Catone (RM), Italy \label{oa-roma}%29
\and Department of Astronomy, University of Geneva, ch. d'Ecogia 16, 1290 Versoix, Switzerland \label{geneva}%12
 }
   \date{Received **; accepted **}

  \abstract
{
We use the final data from the VIPERS redshift survey to extract an unparalleled sample of more than 2000
massive $\cal{M}\ge$10$^{11}$\,M$_{\odot}$ passive galaxies (MPGs) at redshift $0.5 \leq z \leq 1.0$, based on their NUV$rK$ colours. This enables us to investigate how the population of these objects was built up over cosmic time. We find that the evolution of the number density depends on the galaxy mean surface stellar mass density, $\Sigma$. In particular, dense ($\Sigma \geq$ 2000\,M$_{\odot}$pc$^{-2}$) MPGs show a constant comoving number density over this redshift range, whilst this increases by a factor $\sim$ 4 for the least dense objects, defined as having $\Sigma <$ 1000\,M$_{\odot}$pc$^{-2}$. We estimate stellar ages for the MPG population both fitting the Spectral Energy Distribution (SED) and through the D4000$_{n}$ index, obtaining results in good agreement. Our findings are consistent with passive ageing of the stellar content of dense MPGs. We show that at any redshift the less dense MPGs are younger than dense ones and that their stellar populations evolve at a slower rate than predicted by passive evolution. This points to a scenario in which  the overall population of MPGs was built up over the cosmic time by  continuous addition of less dense galaxies: on top of an initial population of dense objects that passively evolves, new, larger, and younger MPGs continuously join the population at later epochs. Finally, we demonstrate that the observed increase in the number density of MPGs is totally accounted for by the observed decrease in the number density of correspondingly massive star forming galaxies
(i.e. all the non-passive $\cal{M}\ge$10$^{11}$\,M$_{\odot}$ objects). Such systems observed at $z\simeq1$ in VIPERS, therefore, represent the most plausible progenitors of the subsequent emerging class of larger MPGs.}
\titlerunning {}
   \authorrunning {Gargiulo et al.}

   \maketitle
%
%________________________________________________________________
\section{Introduction}

Deep imaging and dynamical studies of passive galaxies (PGs, $\cal{M} \gtrsim$ 10$^{10}$\,$\rm{M_\sun}$) have shown that these objects are already in place at $z\sim2$ and, on average, their sizes are a factor $\sim$ 2-4 smaller than comparable galaxies at the current epoch \citep{daddi05,trujillo06, longhetti07, vandokkum08, cimatti08, vanderwel08, bezanson09, cassata11, belli14}. In fact, as example, high-$z$ PGs with $\cal{M} \sim$ 10$^{11}$\,M$_{\odot}$ have, on average, $\it{R_e}$ $\sim$ 1-2\,kpc and mean stellar mass density $\Sigma$ = $\cal{M}$/(2$\pi$$\it{R_e}$$^{2}$) $\gtrsim$ 2000\,M$_{\sun}$pc$^{-2}$, while local PGs with the same stellar mass tend to have $\it{R_e}$ $\sim$ 5\,kpc and $\Sigma \sim 1000$\,M$_{\odot}$pc$^{-2}$ \citep[e.g.][]{shen03, kauffmann03}.

These new data have been widely interpreted within an 'inside-out' scenario according to which  local  PGs accreted their total stellar mass in two phases. During the first phase, a highly dissipative process at high redshift  ($z>$3-5, e.g. a gas-rich merger \citep{hopkins08} or in situ accretion of cold streams \citep{keres05, dekel09} forms a compact passive core. Subsequently, this compact galaxy assembles an external and low-density halo through many dry minor mergers increasing its radius \citep[e.g.][]{naab07, naab09, hilz13, vandokkum08}.

This scenario is supported by the drastic decrease in the number density of compact PGs over time, observed in some studies  \citep[e.g.][]{cassata13, vanderwel14}. However, other analyses have found a very mild decrease \citep[e.g.][]{poggianti13, valentinuzzi10}, or even a constant number density evolution \citep[e.g.][]{damjanov14, damjanov15, saracco10, gargiulo16}. This disagreement casts doubts on the necessity of a size-growth for individual compact PG. In fact, differences in the selection criteria of passive galaxies (morphology vs. colour-colour diagram vs. specific star formation rate [sSFR] cut), in the selection of its dense sub-population (e.g. cut at constant $\it{R_e}$, $\Sigma$, or at 1-2$\sigma$ below the size-mass relation [SMR]), and in the rest-frame waveband used to derive the effective radius, make it difficult to compare results from different studies \citep[or at different cosmic epochs, see e.g.][]{gargiulo16}.

Evidence shows that in conjunction with an increase in the mean radial size of PGs by a factor of  $\sim$ 4 from $z \sim$ 1.5-2 to $z=0$, the number of PGs per unit volume increases by a factor $\sim 10$ \citep[e.g.][]{pozzetti10, ilbert10, brammer11}.
If the number density of high-$z$ dense PGs drastically falls, as observed by some authors, because they increase their size, their contribution to the whole population of local PGs would have to be less then 10$\%$, since high-$z$ PGs are ten times less numerous than local PGs. Consequently, the majority of local PGs would need to have formed through a mechanism other than the inside-out model.
On the other hand, if all local PGs were assembled according to the inside-out model, new compact PGs would need to form continuously at $z <$ 1.5-2 and then increase their sizes in order to sustain the numerical growth of the whole population. This picture could help to explain works that show a mild or null evolution in the number density of compact PGs. However, if this were the case, further evidence would be expected: the age of dense PGs should not evolve significantly over time, since this sub-population would be constantly refreshed by new galaxies. In fact, if the number density of compact PGs is confirmed to evolve slowly, a third hypothesis is also viable: compact PGs might passively age and the increase both in mean $\it{R_e}$ and in the number of PGs over the time could then mostly be due to the fact that galaxies that quench at later epoch are larger \citep[e.g.][]{valentinuzzi10, cimatti08, carollo13}. It is common habit to refer to this last scenario as "progenitor bias" \citep[e.g.][]{franx96}. In this picture the age of the stellar population of compact PGs over cosmic time should be consistent with passive evolution.

Thus, valuable insights into the build up of the PG population can be gained by studying the evolution of the number density {\it and} the age of the stellar population together. 

In this context, massive ($\cal{M} >$ 10$^{11}$\,M$_{\sun}$) PGs (MPGs) deserve particular attention. These systems are expected to evolve mainly through dry mergers \citep[e.g.][]{hopkins09, delucia07} and consequently, to experience a stronger size-growth. So far, because MPGs are extremely rare, very few works have studied the combined evolution of the number density and of the age of MPGs as a function of the compactness of the source \citep{carollo13,fagioli16}. \citet{carollo13}, studied quiescent ($\it{sSFR}$ $<$ 10$^{-11}$\,yr$^{-1}$) and elliptical galaxies in the COSMOS field \citep{scoville07}. They found that the number density of quiescent  and elliptical galaxies with $\it{R_e}$ $<$ 2.5\,kpc decreases by about $30 \%$ from $z \sim 1$ to $z \sim 0.2$ and that their U-V colours are consistent with passive evolution. They concluded that the driving mechanism for the average size-growth of the whole population is the appearance at later epochs of larger quiescent galaxies. More recently, \citet[][hereafter F16]{fagioli16} analysed the spectroscopic properties of  $\sim$ 500 MPGs (defined as galaxies with absent or very weak emission lines and no MIPS detections) at 0.2 $< z <$ 0.8 in the zCOSMOS-bright 20K catalog \citep{lilly07}. From the analysis of stacked spectra of small and large MPGs, they dated the stellar content of these groups and found that the two sub-populations have similar ages. The authors concluded that, in this mass regime, the size growth of individual galaxies through dry mergers is the most likely explanation for the increase in the mean effective radius of the whole population.  A recent analysis by \citet{zahid16} on the physical properties of compact post starburst galaxies at 0.2 $< z <$ 0.8 with $\cal{M} >$ 10$^{11}$\,M$_{\odot}$ provides new insights. On the basis both of their number density and of their ages, which have been found to be $<$ 1Gyr, the authors suggest that this class of objects are the progenitors of compact quiescent galaxies. They conclude that a substantial fraction of dense quiescent galaxies at $z <$ 0.8 are newly formed.

Despite the efforts and improvements of recent years, the overall picture is far from clear. What is still missing is a homogeneous analysis of the number density and stellar population age at z $\sim$ 0.8, i.e. over the redshift range where less compact MPGs start to dominate the Universe \citep[e.g.][]{cassata10, cassata13}. For this goal to be achieved the following three requirements must be met: significant samples of MPGs at different redshifts, in order to have good statistics in each redshift bin and $\Sigma$ bin; a large volume to minimise the effect of the cosmic variance  (which is known to affect the COSMOS field); robust estimates of the ages of stellar populations.

The VIMOS Public Extragalactic Redshift Survey (VIPERS) represents an ideal benchmark for this kind of study. Despite the fact galaxies with $\cal{M}>$10$^{11}$\,M$_{\odot}$ are rare, the wide area ($\sim$ 16 effective deg$^{2}$ over the  W1 and W4 CFHTLS fields, more details in Sect. 2) and high sampling rate of the  survey ($\sim$ 40$\%$) result in a sample of $\sim$ 2000 MPGs with spectroscopic redshifts over the redshift range 0.5$ \leq z \leq$ 1.0 (see Sect.  2). The unprecedented quality of statistics over this redshift range allow us to study the evolution of the number density of MPGs as a function of $\Sigma$ (see Sect. 4). Using D4000$_{n}$ as a spectroscopic diagnostic and by fitting the spectral energy distribution (SED) we constrained the age of the stellar population of MPGs as a function both of $z$ and $\Sigma$ (Sect. 5). This information, together with the evolution of the number density, allowed us to put constraints on the mass accretion scenarios and on the origin of new MPGs (Sect. 6). We summarize all of our results in Sect. 7. Throughout the paper we adopt the Chabrier (2003) initial mass function and a flat cosmology with $\Omega_{M}$ = 0.3 and H$_{0}$ = 70 km\,s$^{-1}$\,Mpc$^{-1}$. Magnitudes are in the AB system. Effective radii are circularized.

\section{Data}
\subsection{The VIPERS project}

In this work we analyse a beta version of the final public data release of VIPERS. The data set used here is almost identical to the publicly released PDR-2 catalogue \citep{scodeggio16}, with the exception of a small sub-set of marginal redshifts (mostly at $z>1.2$), which were revised close to the release date. This has no effect on the analysis presented here.  VIPERS has measured redshifts for 89128 galaxies to $i_{AB}=22.5$, distributed over a total area of 23.5 deg$^{2}$. This reduces to an effective area of 16.3 deg$^{2}$ once detector gaps and masked areas such as those with bright stars are accounted for. Spectroscopic targets were selected in the W1 and W4 fields of the Canada-France-Hawaii Telescope (CFHT) Legacy Survey Wide, applying a robust colour-colour pre-selection in the \textit{ugri} plane, to identify galaxies at $z >$ 0.5. Spectroscopic observations were carried out with VIMOS at the VLT using the low-resolution red grism (R = 220), which covers the wavelength range 5500 - 9500\,\smash{\AA}. The {\it rms} error of the measured redshifts has been estimated to be $\sigma_{z}(z)$ = 0.00054$\times$(1+$z$) \citep{scodeggio16}. In the same paper, a complete description of the PDR-2 data release can be found, while more information on the original survey design and data reduction procedures are found in \citet{guzzo14} and \citet{garilli14}.

\subsubsection{Physical and structural parameters of VIPERS galaxies}

For each galaxy of the VIPERS spectroscopic sample, physical properties such as the multi-band luminosity and the total stellar mass were derived through the fit of the SED. The photometric multi-wavelength coverage combines improved $ugriz$-bands photometry based on the T0007 release of the CFHTLS\footnote{http://www.cfht.hawaii.edu/Science/CFHTLS/} and $Ks$-band observations from the VIPERS Multi-Lambda Survey (VIPERS-MLS\footnote{http://cesam.lam.fr/vipers-mls/}, Moutard et al. \citeyear{moutard16}) or, when available, from the VISTA Deep Extragalactic Observations (VIDEO; Jarvis et al. \citeyear{jarvis13}) survey \citep[more details in][]{davidzon13,davidzon16,moutard16a}. 

The fit was performed with the Hyperzmass software \citep{bolzonella00}. The template libraries adopted in the fit procedure are fully described in \citet{davidzon13} and are based on \citet{bruzual03} models, with exponentially declining star formation history ($\propto$ $e^{-t/\tau}$, where $\tau$ is the time scale of the star formation), $\tau$ in the range [0.1-30]\,Gyr, sub-solar (0.2Z$_{\odot}$) and solar metallicity,  and the \citet{chabrier03} IMF. 
Dust extinction was considered according to two different prescriptions 
 \citep{calzetti00, prevot84}.

The structural parameters (effective radius $\it{R_e}$, Sersic index $n$) 
for $\sim$85$\%$ of the VIPERS sources were derived with GALFIT \citep{peng02} fitting the $i$-band CHFTLS-Wide images with a 2D-psf convolved Sersic profile \citep{krywult16}. The CFHTLS public images have a pixel-scale of $\sim$0.187$''$/px and the full width at half maximum (FWHM) of point-like sources varies from $\sim$0.5$''$ to 0.8$''$ in the $i$ band. To control the variability of the PSF over the wide CCD area of CFHTLS images (1$^{\circ} \times 1^{\circ}$) \citet{krywult16} selected a set of $\sim$ 2000 stars uniformly distributed over each field and modeled their profiles using a 2D Chebychev approximation of the elliptical Moffat function. This approach allowed the PSF to be successfully described over $\sim$ 95$\%$ of the whole VIPERS area. Only regions without bright and unsaturated stars, or at the edge of the images (hereafter bad PSF regions) were excluded. Since these regions are well defined, we have removed them from our analysis. Taking this into consideration the effective final area we used in this analysis is $\sim$ 14 deg$^{2}$.

\citet{krywult16} tested the reliability of $\it{R_e}$ and $n$ derived
from ground-based CFHTLS images. They found that the typical effective galaxy radius is recovered  within 4.4$\%$ and 12$\%$ for 68$\%$ and 95$\%$ of the total sample respectively. We refer to the their paper for further details.

\subsection{The sample of massive passive galaxies}

From the VIPERS spectroscopic catalogue we selected all galaxies with highly accurate redshift measurements, i.e. with quality flag 2 $\leqslant zflag \leqslant$ 9.5 (a confidence level $>$ 95$\%$; 75479 galaxies in the W1+W4 field). We defined galaxies as passive on the basis of their location in the rest-frame colour NUV-$r$ vs. $r$-$K$ diagram, which is a powerful alternative to the UVJ diagram \citep{williams09} to properly identify passive galaxies \citep[for more details see][]{arnouts13, davidzon16, moutard16, moutard16a}.  Following \citet{davidzon16}, we defined quiescent galaxies as those satisfying the following conditions:
\begin{eqnarray}
{\rm NUV} - {\rm r} &>& 3.75, \\ 
 {\rm NUV} - {\rm r} &>& 1.37 \times ({\rm r} - {\rm K}) + 3.2, \\ 
 {\rm r} - {\rm K} &<& 1.3.
\end{eqnarray}
Among all galaxies satisfying these three conditions (7606 in W1 and 3939 in W4, respectively) $\sim$95$\%$ have $\it{sSFR}$ $<$ 10$^{-11}$\,yr$^{-1}$.  From the quiescent population we selected the sub-sample with $\cal{M}$ $\geq$10$^{11}$\,M$_{\sun}$ (hereafter MPGs; 1905 galaxies in the W1 field, and 902 in W4) which is complete up to $z =1.0$ \citep[see e.g.][]{davidzon13,fritz14, davidzon16}. Below $z = 0.5$, VIPERS is highly incomplete due to the \textit{ugri} colour cuts imposed in the selection of spectroscopic targets. Thus the following analysis is limited to the redshift range 0.5 $\leqslant z \leqslant$ 1.0. Finally, we excluded from our analysis galaxies that are in bad PSF regions (see Sect. 2.1). These further cuts leaves us with a sample of 2022 MPGs. 

\begin{figure*}
  \begin{center}
	\includegraphics[angle=0,width=18.9cm, height=5.4cm]{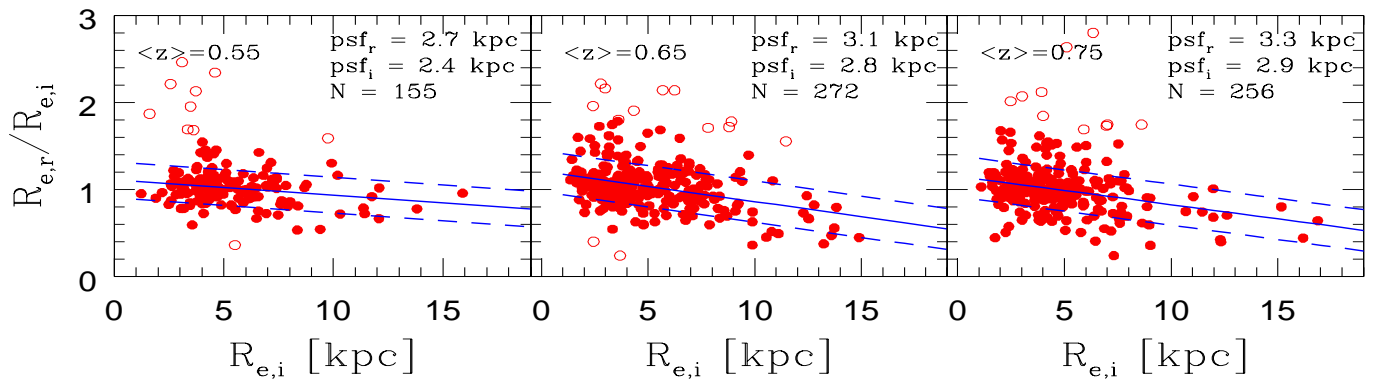} 
	\caption{The ratio between the effective radius in $r$ band ($\it{R_{e,r}}$) and the effective radius in $i$ band ($\it{R_{e,i}}$) as a function of $\it{R_{e,i}}$ for MPGs in three bins of redshift as indicated in the top left corner of each plot. In the three panels, the solid blue lines are the best fit relations derived with a sigma-clipping algorithm. Dashed blue lines set the 1$\sigma$ deviation. The typical dimension of half of the PSF-FWHM of both $i$-band and $r$-band images is indicated in the top right corner of each plot with also the number of objects. Open red circles are galaxies at $>$ 3$\sigma$ from this fit.}
	\label{radius}
  \end{center}
\end{figure*}

For each of these galaxies we need to derive its mean stellar mass density $\Sigma$. In the redshift range $0.5 \leqslant z \leqslant 1.0$, the $i$-band filter covers the spectral region across the 4000$\smash{\AA}$ break. In particular, it samples the $\sim$ V-band (5000\smash{\AA}\,) rest-frame at $z = 0.5$ and  the $\sim$ U-band (3500\smash{\AA}\,) rest-frame at $z = 1.0$. Given the presence of radial colour variations in passive galaxies both at high and low redshift \citep[see e.g.][]{labarbera09, saglia10, gargiulo12, guo11}, the variation of the rest-frame band used to derive $\it{R_e}$ can induce a spurious trend in the evolution of the radius with redshift. To quantify this bias, the structural parameters for the whole W4 field (and for $\sim$40$\%$ of the W1 field) have been derived in the $r$-band (FWHM $\sim$0.8$''$, $\sim$ U-band rest-frame at $z \lesssim$ 0.8). Figure \ref{radius} shows the ratio between the $\it{R_e}$ of MPGs in the $r$ and $i$ band ($\it{R_{e,r}}$/$\it{R_{e,i}}$) as a function of their $\it{R_{e,i}}$ at $z < 0.8$. For a rigorous analysis, we derived the relations in three finer redshift bins (0.5 $\leqslant z <$ 0.6, 0.6 $\leqslant z <$ 0.7, 0.7 $\leqslant z <$ 0.8). The general trend, as derived from the best-fit relations, is that for MPGs $\it{R_{e,i}} < \it{R_{e,r}}$. In other words, the internal regions are redder than the outskirts. For the smallest galaxies the difference between the two radii can be up to $\sim$20$\%$. We note that a portion of MPGs (especially at large $\it{R_e}$) has $\it{R_{e,i}} > \it{R_{e,r}}$, suggesting the presence a of blue core. However, a detailed study on the internal colour variation in MPGs is beyond the scope of this paper.
Given the evidence presented in Fig. \ref{radius}, in the derivation of $\Sigma$ we referred to the effective radius in the $r$ band for MPGs at $z < 0.8$, and to $\it{R_e}$ in the $i$ band for those at $z \geq 0.8$. By doing this, we are able to approximately sample the same U band rest-frame over the whole redshift range we probe. For those MPGs at $z < 0.8$ without an $\it{R_{e,r}}$ estimate (mostly in the W1 field), we estimated $\it{R_{e,i}}$ using the relations of Fig. \ref{radius}. We checked that the addition of galaxies with $\it{R_{e,r}}$ derived from $\it{R_{e,i}}$ does not change the $\Sigma$ distribution of the sample of MPGs at $z < 0.8$.

\begin{figure}
  \begin{center}
   	\includegraphics[angle=0,width=8.9cm]{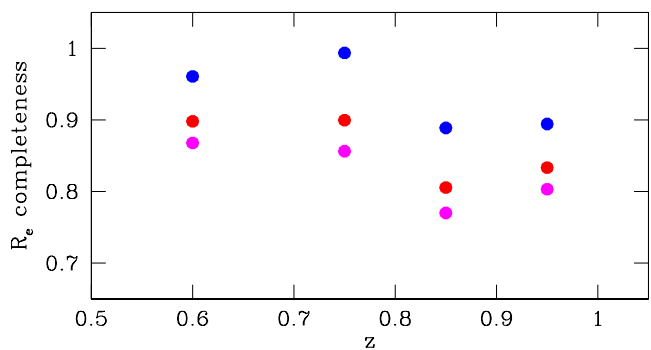} \\
	\caption{Fraction of MPGs with available $\it{R_e}$ in the W1 field (magenta filled points), in the W4 field (blue filled points) as a function of $z$. At $z <$ 0.8, MPGs with $\it{R_e}$ derived from $i$-band images are included. Red filled points indicate the completeness for the W1+W4 field.}
	\label{stsr}
  \end{center}
\end{figure}

In Figure \ref{stsr} we show the fraction of MPGs with available and reliable structural parameters for W1 and W4 fields (magenta and blue points, respectively) and for the total area (red points) as a function of redshift.

Overall, $\sim$85$\%$ of MPGs in the two fields have a reliable $\it{R_e}$. For $\sim$15$\%$ of MPGs structural parameters are not available since in some cases we are unable to fit the surface brightness profile. This is either because the algorithm does not converge, or the best-fit values of n $<$ 0.2 are unphysical \citep[see][]{krywult16}. Once excluded these objects, the final sample of MPGs with reliable $\it{R_e}$ (hence $\Sigma$) in the redshift range 0.5 -1.0 is composed of 1758 galaxies.

\section{The size mass relation of MPGs in VIPERS for $0.5 < z < 1.0$}

\begin{figure}
  \begin{center}
   	\includegraphics[angle=0,width=9.1cm]{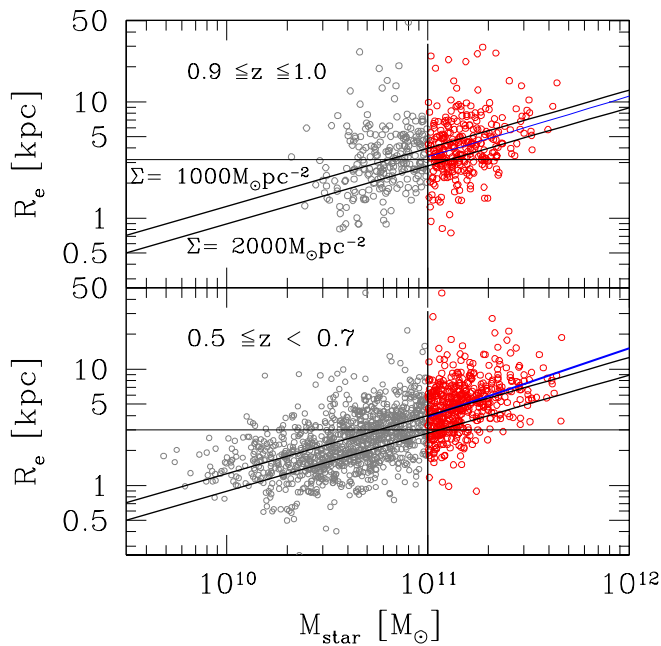} \\
	\caption{The size-mass relation of passive VIPERS galaxies (grey open + red open points) in the two extreme redshift bins of our analysis (0.5 $\leqslant z <$ 0.7 lower panel, and 0.9 $\leqslant z \leqslant $ 1.0 upper panel). Blue open points are the MPGs analyzed in this paper (vertical lines indicate the mass cut). 
Horizontal lines indicate half of the FWHM at that redshift, while magenta lines are the best fit of the size-mass relation
of MPGs. Black diagonal lines are lines of constant $\Sigma$  (2000 and 1000\,M$_{\odot}$pc$^{-2}$).}
	\label{sizemass}
  \end{center}
\end{figure}

In Fig. \ref{sizemass} we show the SMR for the VIPERS MPGs in the lowest and highest redshift bin of our sample, i.e. 0.5 $\leqslant z <$ 0.7 and 0.9 $\leqslant z \leqslant$ 1.0. We fitted a size-mass relation of MPGs of the form $\log$ $\it{R_e}$ = $\alpha$ $\log$ ($\cal{M}$/10$^{11}$) + $\beta$  adopting an ordinary least squares fit without take into account errors on $\it{R_e}$.
The best-fit results are reported in Table \ref{fit}. We add also the result for the intermediate bin (0.7 $\leq z <$ 0.9).
\begin{table}
\caption{The best-fit values ($\alpha$, $\beta$) of the size mass relation $\log$ $\it{R_e}$ = $\alpha$ $\log$ ($\cal{M}$/10$^{11}$) + $\beta$ of VIPERS MPGs. $\it{R_{e,11}}$ indicates the value of $\it{R_e}$ at $\cal{M}$ = 10$^{11}$\,M$_{\odot}$ as predicted by the best-fit relation.}
  \begin{center}
  \begin{tabular}{cccc}
  \hline
  \hline
  z &  $\alpha$ & $\beta$ & R$_{e,11}$[kpc]\\
  0.5 $\leqslant z <$ 0.7 & 0.59$\pm$0.07 & 0.60$\pm$0.01 & 3.9\\
  0.7 $\leqslant z <$ 0.9 & 0.70$\pm$0.08 & 0.53$\pm$0.02 & 3.4\\
%  0.8 $\leqslant z <$ 0.9 & 0.81$\pm$0.09 & 0.49$\pm$0.02 & 3.3\\
  0.9 $\leqslant z \leqslant$ 1.0 & 0.52$\pm$0.10 & 0.52$\pm$0.02 & 3.3\\
\hline
\hline
  \end{tabular} 
    \label{fit}
  \end{center}
\end{table}
In agreement with previous studies \citep[e.g.][]{damjanov11, vanderwel14}, we find that at 1$\sigma$ there is almost no evolution of the slope of the SMR with time. However, there is an offset among the zero-points. On average, at fixed stellar mass, MPGs at $\langle z \rangle = 0.6$ have $\it{R_e}$ $\sim$ 1.25 larger than those at $z = 1.0$. This increase shows that the growth in the mean $\it{R_e}$ of the passive population is gradual and continuous, extending out to $z < 1$. The increase in $\it{R_e}$ is in agreement with what has been found by other authors. \citet{damjanov11}, for a sample of erly-type galaxies at 0.2 $< z <$ 2.7, found $\langle R_{e} \rangle$ $\propto$ (1+$z$)$^{-1.62\pm0.34}$, independently of  the stellar mass of the galaxy, i.e. an increase by a factor $\sim$1.4$\pm$0.1 in the redshift range 0.6 - 1.0. Similarly, \citet{williams10} found $\langle R_{e} \rangle \propto$ (1+$z$)$^{-1.3}$ for passive galaxies (UVJ selected) with $\cal{M} >$ 10$^{11}$\,M$_{\odot}$ which results in an increase of a factor $\sim$1.3 in our redshift range. From the analysis of passive galaxies (UVJ colour selected) with $\cal{M} >$ 2$\times$10$^{10}$\,M$_{\odot}$, \citet{vanderwel14} found $\langle R_{e}\rangle \propto$ (1+$z$)$^{-1.48}$ over the redshift range $0 < z < 3$, still in fair agreement with our results. 

\begin{figure}
  \begin{center}
   	\includegraphics[angle=0,width=9.0cm]{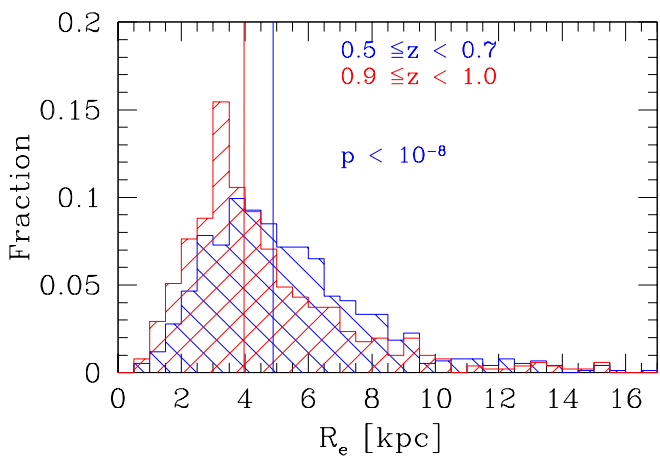} \\
	\caption{The distribution of the effective radius for MPGs in our lowest (blue histogram) and highest (red histogram) redshift bins. The solid lines indicate the median values of the two distributions. The probability p that the two distributions are extracted from the same parent sample are reported in the label.}
	\label{sizedist}
  \end{center}
\end{figure}

In Fig. \ref{sizedist} we directly show the distribution of $\it{R_e}$ of MPGs in the same redshift bins as Fig. \ref{sizemass}. Although the two distributions are overall different (KS test probability $<$ 10$^{-8}$), the Figure shows that they cover the same range of $\it{R_e}$. This suggests that the population of compact MPGs does not totally disappear with cosmic time. In the next section we will quantify this qualitative trend by estimating the comoving number density of MPGs with different $\Sigma$.

\section{The number density of MPGs in VIPERS as a function
of redshift and mean stellar mass density}

In order to derive the number density of the MPGs a function of $\Sigma$ we subdivided our sample into a ``high-$\Sigma$ sample'', consisting of galaxies with $\Sigma \geq$ 2000\,M$_{\odot}$pc$^{-2}$, an ``intermediate-$\Sigma$ sample'', consisting of galaxies with 1000 $< \Sigma \leqslant$ 2000\,M$_{\odot}$pc$^{-2}$, and a ``low-$\Sigma$ sample'' consisting of galaxies with $\Sigma <$ 1000\,M$_{\odot}$pc$^{-2}$. The choice of a cut at 2000\,M$_{\odot}$pc$^{-2}$ was determined by the characteristics of the CFHTLS images. In Fig. \ref{sizemass}, the horizontal solid lines indicate half of the FWHM of the PSF. We note that all the  MPGs with $\it{R_e}$ below half the FWHM of the images, at any redshift, have $\Sigma \gtrsim$ 2000\,M$_{\odot}$pc$^{-2}$. Adopting this cut ensures that all galaxies with radii lower than the resolution of the image (i.e. galaxies for which $\it{R_e}$ could be overestimated), are in the same $\Sigma$ bin and do not spuriously contaminate other bins.  We stress that this is a conservative choice. Effective radii are derived by deconvolving the data for the real PSF of the images. At the typical S/N of our galaxies, measurements of $\it{R_e}$ lower than the image resolution are robust as shown in the Appendix of \citet{krywult16}.

In Tab. \ref{tabnd}, we list the total number of galaxies in each bin of stellar mass density, and for each redshift bin. Their sum is smaller than the total number of galaxies (first column in Tab. \ref{tabnd}) since for a fraction of galaxies we do not have a reliable estimate of their structural parameters (see Sect. 2.2). 
\begin{table}
\caption{The total number of massive quiescent galaxies and of the high-$\Sigma$ sample, intermediate-$\Sigma$ sample, and low-$\Sigma$ sample in the four redshift bins 0.5 $ \leqslant z <$ 0.7, 0.7 $ < z \leqslant$ 0.8, 0.8 $ < z \leqslant$ 0.9, 0.9 $ < z \leqslant$ 1.0. At each redshift the sum of the high+intermediate+low $\Sigma$ galaxies does not return the total number of objects since not all the MPGs have a reliable $\it{R_e}$.}

  \begin{center}
\begin{tabular}{ccccc}
\hline
\hline
 z & N$_{tot}$ & N$_{high-\Sigma}$ & N$_{int-\Sigma}$ & N$_{low-\Sigma}$ \\
\hline
\hline
0.5 - 0.7 & 782 & 165 & 185 & 352\\
0.7 - 0.8 & 482 & 144 & 128 & 163\\
0.8 - 0.9 & 386 & 103 & 92  & 116\\
0.9 - 1.0 & 372 & 112 & 97  & 101\\
%1.0 - 1.1 & 193 &  89 & 47  &  57 \\
\hline
\hline
\end{tabular}
  \label{tabnd}
\end{center}
\end{table}

	The VIPERS final sample (and hence our final sample of MPGs) suffers from three sources of incompleteness. These are the Target Sampling Rate (TSR), the Success Sampling Rate (SSR), and the Colour Sampling Rate (CSR). The TSR is given by the ratio of galaxies effectively observed and targeted galaxies. The SSR is the fraction of spectroscopically observed galaxies with a redshift measurement. The CSR takes into account the completeness due to the colour selection of the survey. These statistical weights (hereafter \textit{TSR}(i), \textit{SSR}(i), \textit{CSR}(i)) depend on the magnitude of the galaxy, on its redshift, colour, and angular position. They have been derived for each galaxy in the full VIPERS sample \citep[for a detailed description of their derivation see][]{garilli14, scodeggio16}.  In the derivation of the number density, we weighted each MPG $i$ in our sample by the quantity \textit{w(i)} = 1/(\textit{TSR(i)}*\textit{SSR(i)}*\textit{CSR(i)}). 

Beside these sources of incompleteness, a fraction of MPGs is lacking of reliable structural parameters. In Fig. \ref{stsr} we show this fraction as a function of redshift. We checked whether the galaxies without structural parameters belong mainly to a sub-population of galaxies of a given $\Sigma$. To address this issue, we compared the fraction of high-, intermediate- and low-$\Sigma$ MPGs in different redshift bins (which have different levels of completeness). We did not find any significant variation between redshift bins. Given that there is no dependence between the lack of structural parameters and $\Sigma$, we corrected the number densities of MPGs for this source of incompleteness, using the values in Fig. \ref{stsr}.

\subsection{The number densities of MPGs in VIPERS field
as a function of z and $\Sigma$}

\begin{figure*}
  \begin{center}
   	\includegraphics[angle=0,width=18.5cm]{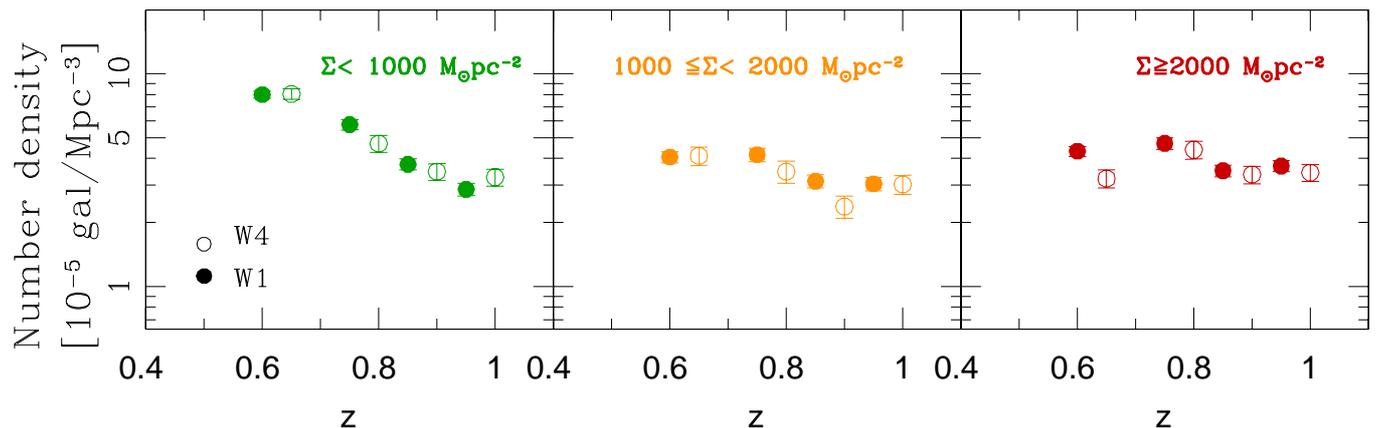} \\
	\caption{The number densities for massive passive galaxies with $\Sigma \geq$ 2000\,M$_{\odot}$pc$^{-2}$ (right panel, dark red symbols), 1000 $ \leq \Sigma <$ 2000\,M$_{\odot}$pc$^{-2}$ (central panel, orange symbols), and  $\Sigma <$ 1000\,M$_{\odot}$pc$^{-2}$ (left panel, green symbols), for the W1 field (filled circles) and W4 one (open circles). Number densities for the W4 field are shifted in redshift just to visualize them better. The error bars correspond to 1$\sigma$.}
	\label{nd3p}
  \end{center}
\end{figure*}

In Fig. \ref{nd3p} we show the number density of MPGs as a function of redshift and mean stellar mass density, both in the W1 and W4 fields.
Error bars were derived taking into consideration Poisson fluctuations, and the uncertainties on the $\it{R_e}$ estimates. 
To consider this last source of uncertainty, we computed the standard deviation $\sigma$($z$, $\Sigma$) over 100 number density estimates obtained by replacing for each galaxy the effective radius $\it{R_e}$ with a value randomly drawn from a Gaussian distribution with mean value $\it{R_e}$ and standard deviation the typical error on $\it{R_e}$ (i.e. 0.05$\it{R_e}$, see Sect 2.1).

\begin{figure}
 \begin{center}
  	\includegraphics[angle=0,width=8.8cm]{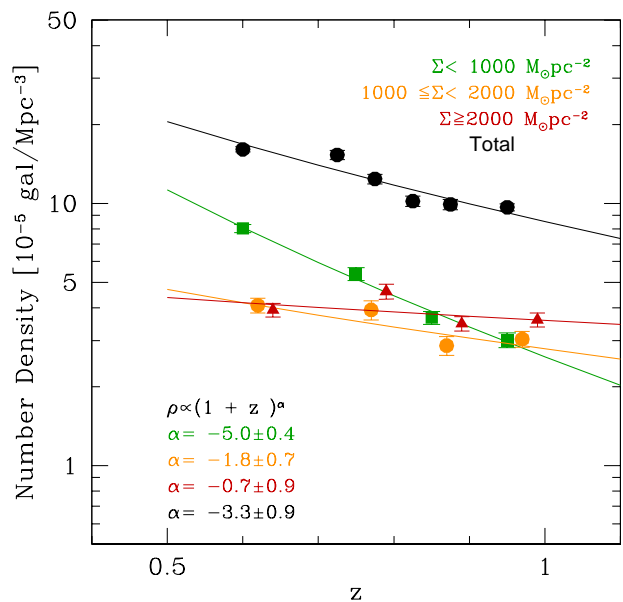} \\
	\caption{The number density of MPGs with different mean stellar mass density in VIPERS field as a function of the redshift. Dark red triangles refer to high-$\Sigma$ MPGs, orange circles refer to intermediate-$\Sigma$ MPGs, and green squares to low-$\Sigma$ MPGs. Black circles show the number density for the total sample of MPGs. Number densities for the three sub-populations are shifted in redshift just to visualize them better. The error bars correspond to 1$\sigma$. We fitted the data with a power law and solid lines are the best-fit relations.}
	\label{numdentot}
 \end{center}
\end{figure}

The estimates of the number density in the W1 and W4 fields are in very good agreement, indicating that the wide area of VIPERS reduces the effect of cosmic variance even for the most massive galaxy sample. In Fig. \ref{numdentot} we report the number density of MPGs as a function of $\Sigma$ (plus the number density for the whole population of MPGs) in the total VIPERS area (W1 + W4). Error bars are derived as described above. 
The evolution of the number density of the whole population of MPGs is well fitted by a function $\rho \propto (1+$z$)^\alpha$, with $\alpha$ = -3.3$\pm$0.9. In the $\sim$ 2.5 Gyr from $z = 1.0$ to $z = 0.5$, the number density of MPGs increases by a factor $\sim$ 2.5. Similarly, we fitted the number density evolution of the three sub-populations with a power law, and found $\alpha$ = -5.0$\pm$0.4 for low-$\Sigma$ MPGs, $\alpha$ = -1.8$\pm$0.7 for intermediate-$\Sigma$ MPGs and $\alpha$ = -0.7$\pm$0.9 for high-$\Sigma$ MPGs. We find that the evolution of the number density strongly depends on the mean stellar mass density of the system. In particular the lower the mean stellar mass density, the faster the evolution. The number density of the densest MPGs is approximately constant over the whole redshift range, while the number density of less dense MPGs constantly increases with time by a factor $\sim$4.

Figure \ref{numdentot} shows that we are looking at  a crucial moment in the build up of the MPG population, i.e. when less compact galaxies, which constitute the bulk of the local MPG population, start to dominate. In fact, at $z > 0.8$, the population of MPGs is composed in equal parts of high, intermediate and low-$\Sigma$ galaxies. At lower redshift, the contribution of low-$\Sigma$ MPGs steadily increases. At $z = 0.5$ compact quiescent galaxies account for just $\sim$15$\%$ of the whole population. Less compact systems, on the other hand, account for more than half of the whole population. This result is in agreement with the analysis by \citet{cassata11} who found that normal (i.e. less dense) passive ($\it{sSFR}$ < 10$^{-11}$\,yr$^{-1}$) and elliptical galaxies with 10$^{10} < $ $\cal{M} \lesssim$ 10$^{11.5}$\,M$_{\odot}$ start to become the dominant sub-population at $z \sim$ 0.9. 
 
Our results are in good agreement also with \citet{carollo13} who found that the number density of passive ($\it{sSFR}$ $<$ 10$^{-11}$\,yr$^{-1}$) and elliptical galaxies with $\cal{M} >$10$^{11}$\,M$_{\odot}$ and $\it{R_e} <$ 2.5\,kpc decreases by $\sim$ 30$\%$ in the 5\,Gyr between $z \sim$ 1 and $z \sim$ 0.2. \citet{damjanov15} found an almost constant number density over the redshift range 0.2$ < z <$0.8 for colour selected passive compact galaxies with $\cal{M} >$ 8$\times$10$^{10}$\,M$_{\odot}$. \citet{gargiulo16} found $\rho(z)\propto(1 + z)^{0.3\pm0.8}$ for a sample of morphologically selected elliptical dense galaxies ($\Sigma \geqslant$ 2500\,M$_{\odot}$pc$^{-2}$) with $\cal{M} >$10$^{11}$\,M$_{\odot}$ in the redshift range 0 $< z < 1.6$, consistent with our results. \citet{cassata13} found a very mild decrease of compact galaxies (1$\sigma$ below the local SMR), and a mild increase in the number density of normal galaxies (consistent at 1$\sigma$ with the local SMR) from $z \sim$ 1 to $z \sim$ 0.5, in qualitative agreement with our results. However, they found that the number density of ultra-compact galaxies (0.4dex smaller than local SDSS counterparts of the same mass) dramatically decreases. A detailed comparison with our results is not possible given the different selection of the samples. Nonetheless, we verified that the constant trend for high-$\Sigma$ MPGs we show in Fig. \ref{numdentot} is not related to the adopted cut $\Sigma$ = 2000\,M$_{\odot}$pc$^{-2}$. We estimated the number density for the sub-sample of MPGs with $\Sigma >$  3000\,M$_{\odot}$pc$^{-2}$ and $\Sigma >$  4000\,M$_{\odot}$pc$^{-2}$ and found $\alpha$ =  -0.5$\pm$1.4 and $\alpha$ = -1.3$\pm$1.5. These results are consistent at 1$\sigma$ with the results we found for the high-$\Sigma$ sub-population.

If the global population of dense passive galaxies were to evolve in size, then in order to maintain a constant number density of the compact sub-population over cosmic time, new dense galaxies would have to appear at lower redshift. In particular, the time scale of the size-growth mechanisms and that of the appearance of new dense massive quiescent galaxies would need to be very similar \citep[e.g.][]{carollo13}. The other possibility is that the dense MPGs passively evolve without changing their structure. One way to discriminate between these two possibilities is to study of the age of 
the stellar population as a function of the mean stellar mass density.

\section{The stellar population age of MPGs as a 
function of the redshift and mean stellar mass density}

For any individual galaxy, we constrained its stellar population age both fitting the photometric SED (age$_{SED}$) and through the D4000$_{n}$ index. The D4000$_{n}$ index \citep{balogh99} is an age sensitive spectral feature \citep{kauffmann03} defined as the ratio between the continuum flux densities in the blue region [3850 - 3950]\smash{\AA}\, and  red region [4000 - 4100]\smash{\AA}\ across the 4000\smash{\AA}\, break:
\begin{equation}
 D4000_n = \frac{(\lambda_2^{\rm blue} - \lambda_1^{\rm blue})\int_{\lambda_1^{\rm red}}^{\lambda_2^{\rm red}}F_{\nu}\,d\lambda}{(\lambda_2^{\rm red} - \lambda_1^{\rm red})\int_{\lambda_1^{\rm blue}}^{\lambda_2^{\rm blue}}F_{\nu}\,d\lambda}.
\end{equation}
A complete description of the D4000$_{n}$ measurements for VIPERS galaxies is presented in \citet{garilli14}.  \citet{siudek16}, using staked spectra, have already investigated the star formation epoch of all of VIPERS passive galaxies through the analysis of both the D4000$_{n}$ and the H$_{\delta}$ Lick index. The authors selected passive galaxies using an evolving cut in the rest-frame U-V colour and found that, over the full analyzed redshift and stellar mass range (0.4 $< z <$ 1.0 and 10 $<$ log($\cal{M}$/M$_{\odot}$) $<$12, respectively), the D4000$_{n}$ index increases with redshift, while H$_{\delta}$ gets lower. Here, instead of looking for trends of D4000$_{n}$ with stellar mass and $z$, we focused our analysis on a given stellar mass bin, and investigate within this bin the trend of D4000$_{n}$ with the $\Sigma$ and $z$.
We selected MPGs with the error on D4000$_{n}$ smaller than 8$\%$ in order to restrict the analysis to galaxies with a highly accurate estimate of D4000$_{n}$. This further criterion reduces the sample by $\sim$10$\%$, but does not alter the distribution in $\Sigma$. Fig. \ref{models} in Appendix A shows how the conversion of the D4000$_{n}$ into a stellar population age depends both on metallicity $Z$ and on the time scale $\tau$ of the star formation of the galaxy (we report the trend in the case of an exponentially declining star formation history). Given the spectral coverage and resolution of our spectra, we can accurately measure D4000$_{n}$ for each individual galaxy but conversely we cannot constrain either $Z$ or the time scale. Thus, instead of assuming some values we adopted the following approach. Starting from the best-fit models of the SED, we derived the mean value of $Age_{\rm SED}$($z$,$\Sigma$), $Z_{\rm SED}$($z$,$\Sigma$), and $\tau_{\rm SED}$($z$,$\Sigma$) for MPGs in each bin of $z$ and $\Sigma$. Using low-resolution (lr) BC03 models (i.e. the same as those used to perform the SED fitting),  we derived the D4000$_{n}$ corresponding to these mean values, D4000$_{n, SED}$($z$,$\Sigma)$, and compared this estimate with the mean value of the distribution of D4000$_{n}$ of all of the MPGs in that bin of $z$ and $\Sigma$. For simplicity we focus this part of our analysis on the two extreme $\Sigma$ sub-populations, the low-$\Sigma$ and high-$\Sigma$ MPGs. Results for intermediate-$\Sigma$ MPGs are in between.

\begin{figure*}
 \begin{center}
 \begin{tabular}{cc}
 	\includegraphics[angle=0,width=8.5cm]{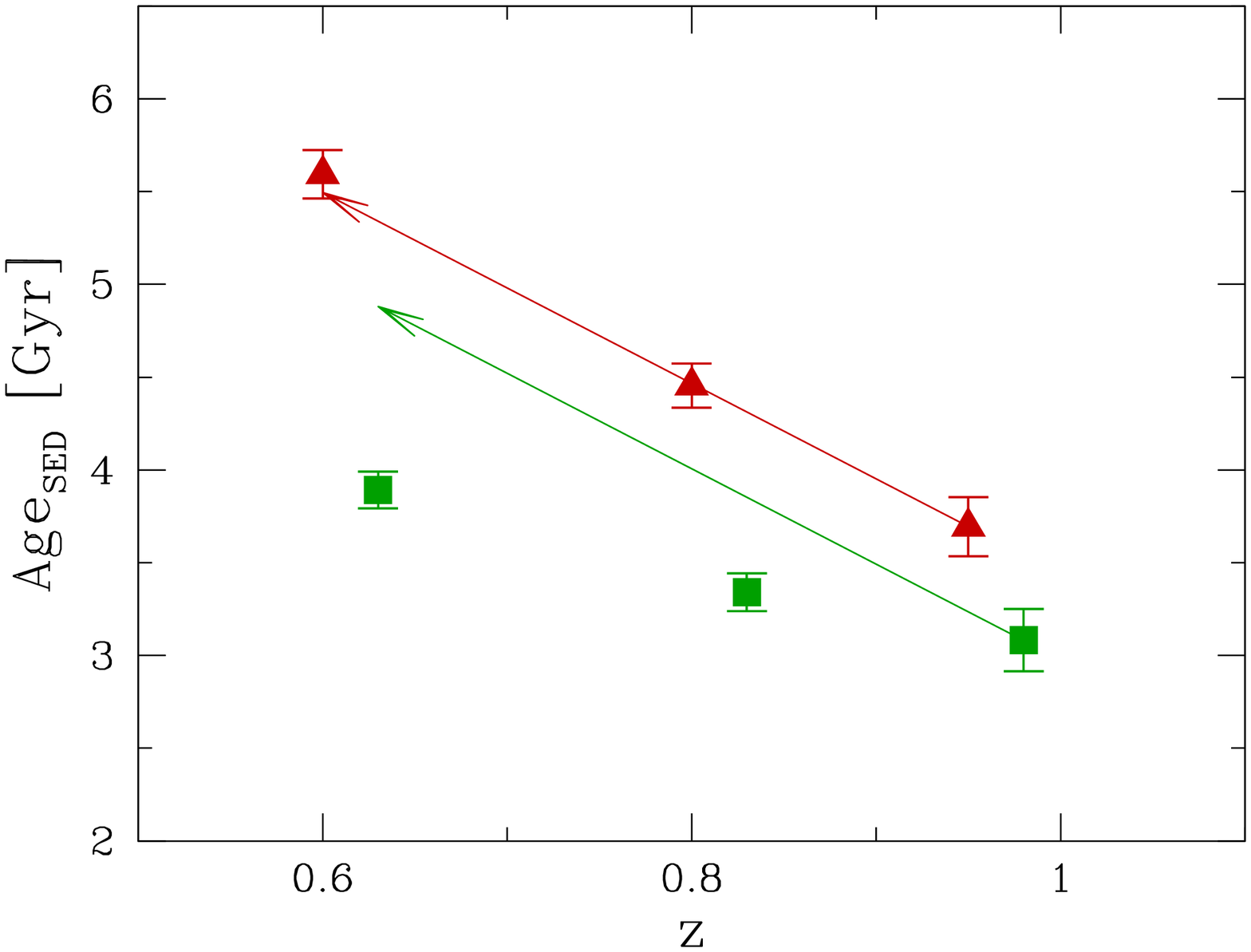} &  
\includegraphics[angle=0,width=8.5cm]{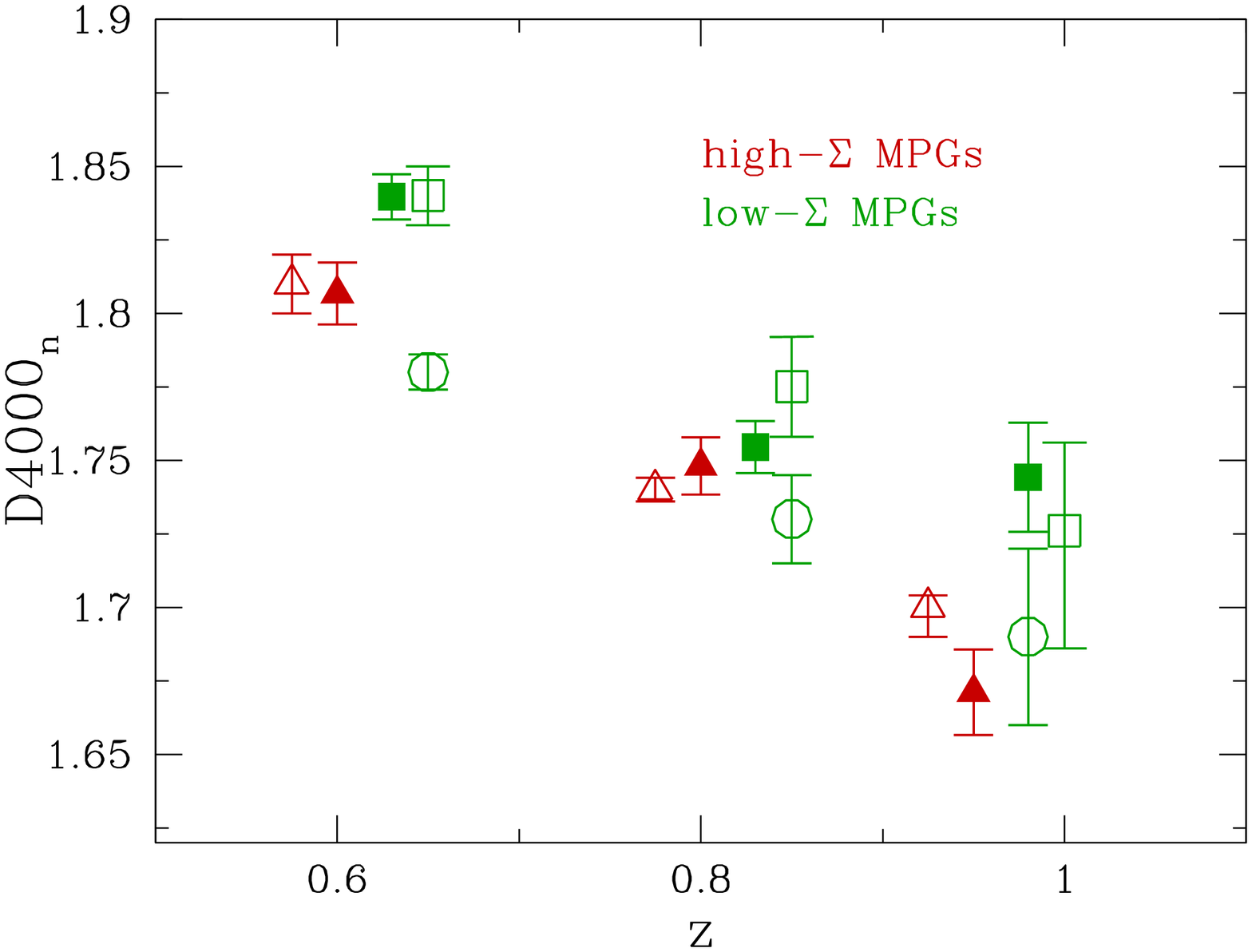}
  \end{tabular}
 	\caption{\textit{Left panel}: Mean stellar population ages of MPGs as derived from the SED fitting as a function of redshift and mean stellar mass density for high- and low-$\Sigma$ MPGs (filled points). The colour and symbols code is the same as in Fig. \ref{numdentot}. Error bars indicate the error on the mean. Points refer to the redshift bins 0.5 $\leqslant z <$ 0.7, 0.7 $\leqslant z <$ 0.9, 0.9 $\leqslant z \leqslant$ 1.0. The arrows track the increase of the ages in case of passive evolution from $z = 0.95$ to $z = 0.6$. \textit{Right panel}: Filled points: the mean values of D4000$_{n}$ directly measured on spectra for MPGs with different $\Sigma$ (symbol/color code the same as in the left panel). Open dark red triangles indicate the values of D4000$_{n}$ of high-$\Sigma$ MPGs derived from a BC03 model with age/$\tau$/$Z$ equal to the mean values derived from the SED fitting of this sub-population (D4000$_{n, SED}$).  Open green squares are the equivalent for low-$\Sigma$ MPGs, once the aperture bias is taken into account. Actually, for high-$\Sigma$ MPGs the slit aperture samples approximately the whole galaxy (see Fig. \ref{ratioevol}), thus  D4000$_{n}$ measured directly from the spectra and  D4000$_{n, SED}$ can be fairly compared. For low-$\Sigma$ MPGs Fig. \ref{ratioevol} shows that the slit samples just the region within 0.5$\it{R_e}$. In the comparison of D4000$_{n}$ and D4000$_{n, SED}$ hence, we have to take into account the effect of colour gradients (see more details in Sect. 5.2). Open green circles indicate the values of D4000$_{n, SED}$ of low-$\Sigma$ without any correction for aperture bias.}
	\label{d4evo}
 \end{center}
\end{figure*}
In Fig. \ref{d4evo} filled points show the mean values of Age$_{\rm SED}$ and the mean value of D4000$_{n}$ (left and right panels, respectively) of low-$\Sigma$ (green filled squares) and high-$\Sigma$ (dark red triangles) MPGs. Error bars indicate the error on the mean. The left panel of Fig. \ref{d4evo} shows that for high-$\Sigma$ MPGs the evolution of the stellar population ages derived from the SED fitting is consistent with a passive evolution of the population. In fact, their Age$_{\rm SED}$ increases by $\sim$2Gyr during the 1.8 Gyr of evolution between $z = 0.95$ and 0.6. At any redshift, the mean value of the time scale of star formation $\tau$, as constrained by the SED fitting, is  $\sim$ 0.4\,Gyr, and the best-fit models have a mean $Z$ $\sim$ 0.5$\pm$0.3Z$_{\sun}$\footnote{We caution that this value refers to the whole galaxy, not just to the central region known to have $Z$ $>$ Z$_{sun}$ \citep[e.g.][]{gallazzi05}. This value is in agreement with the mean metallicity within $\it{R_e}$ of local massive ETGs (0.7$^{+0.15}_{-0.12}$Z$_{\sun}$, see Appendix C).}. Using these constraints on age/$\tau$/$Z$ and BC03 models, we derived D4000$_{n, SED}$. Actually, BC03 models provide the theoretical SED for discrete values of $Z$ (e.g. $Z$ = 0.2Z$_{\odot}$, 0.4Z$_{\odot}$, Z$_{\odot}$). Thus, to derive the D4000$_{n, SED}$ we interpolated the estimates of the two models which  encompass the mean values of $Z$. The resolution of BC03 lr models is 20\smash{\AA} while VIPERS spectra have a resolution of $\sim$ 17\smash{\AA}\, around the 4000\smash{\AA}\, break. Using the high resolution (3\smash{\AA})\, BC03 models we verified that this difference affects the D4000$_{n}$ by $\sim$0.01 (see Appendix A). This is well below the typical error on D4000$_{n}$ and for this reason we did not correct for it. The values of D4000$_{n, SED}$ are shown in the right hand panel of Fig. \ref{d4evo} with open dark red triangles. They are in good agreement with the mean value of D4000$_{n}$ measured on real spectra. We stress that this comparison is meaningful only between the mean values of the distributions and not for any single galaxy. Conventionally SED modeling uses a coarse grid of $Z$/$\tau$, that cannot accurately reproduce the more realistic smooth distribution of metallicity nor time scale of star formation of real galaxies. Nonetheless, although the single values of stellar population parameters could be biased, the mean values are more representative of the truth as suggested by the consistency between D4000$_{n}$ and D4000$_{n, SED}$ in Fig. \ref{d4evo}. The results of Fig. \ref{d4evo} and Fig. \ref{numdentot} show that both the evolution of the number density and of the age of stellar population of dense MPGs, are coherent with passive evolution since $z = 1.0$. 

For what concerns the low-$\Sigma$ MPGs, the left-hand panel of Fig. \ref{d4evo} shows that at any redshift they are systematically younger than high-$\Sigma$ MPGs. In particular their ages increase by just 0.4\,Gyr in the 1.8\,Gyr of time that passes between $z = 0.95$ and 0.6. Before comparing the observed value of D4000$_{n}$ with D4000$_{n, SED}$  it is important to note that the D4000$_{n}$ we used refers to the portion of the galaxy that falls into the 1$''$ slit. Considering the different dimensions of high, intermediate, and low-$\Sigma$ MPGs, the slit samples a different fraction of total light for the three sub-populations.
\begin{figure}
 \begin{center}
  	\includegraphics[angle=0,width=9.0cm]{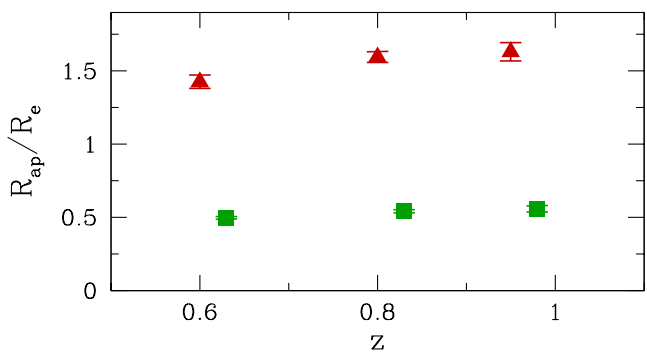} \\
	\caption{The ratio between half of the aperture of the slit and effective radius of the galaxies as a function of the mean 	stellar mass density and z. The colour code is the same as in Fig. \ref{numdentot}.}
	\label{ratioevol}
 \end{center}
\end{figure}
In Fig. \ref{ratioevol}, the points indicate the portion of galaxy covered by the slit in $\it{R_e}$ unit for the high-$\Sigma$ and low-$\Sigma$ sub-populations, as a function of redshift. Almost all of the light from high-$\Sigma$ MPGs is included in the slit, thus the D4000$_{n}$ provides constraints on the age of the whole galaxy, similarly to the SED fitting. For low-$\Sigma$ MPGs, the slit samples the inner 0.5$\it{R_e}$. Given the known presence of metallicity gradients in passive galaxies both in local Universe and at intermediate redshift, this implies that at any redshift i) we cannot directly compare the D4000$_{n}$ of high- and low-$\Sigma$ MPGs, and ii) for low-$\Sigma$ MPGs, we cannot directly compare the results from SED fitting with those derived from the spectral features as we did for high-$\Sigma$ MPGs. Although a direct comparison at fixed redshift is not straightforward, we highlight that the region which is sampled by the slit is approximately  constant over our redshift range (see Fig. \ref{ratioevol}) and this assures us that the evolutionary trends of D4000$_{n}$ are not affected by aperture effects.

Taking advantage of the spatially resolved information on stellar population properties provided by ATLAS$^{3D}$ survey \citep{cappellari11} for a sample of 260 local elliptical galaxies (ETGs), we quantified that for low-$\Sigma$ local massive ETGs the metallicity in the central (r $<$ 0.5$\it{R_e}$, $Z$$_{Re/2}$) region is $\sim$ 20$\%$ greater than the metallicity within $\it{R_e}$ (see Appendix C). At the same time, we found that there is no significant trend with radius of the time scale of the star formation or age\footnote{The population of passive galaxies and elliptical galaxies are not coincident \citep[see e.g.][]{tamburri14, moresco13}. However, at first order, the two populations share the same properties.}. Taking this into consideration, we derived D4000$_{n, SED}$ applying the correction we derived from local Universe to the mean values of $Z$ (note that \citet{gallazzi14} do not find any evolution in the metallicities of massive quiescent galaxies since $z \sim$ 0.8).  Results are shown as open green squares in the right panel of Fig. \ref{d4evo}.  As for high-$\Sigma$ MPGs, they are in good agreement with the D4000$_{n}$ values measured on real spectra. Just to qualitatively show the effect of slit aperture, open circles are the values of D4000$_{n, SED}$ not corrected for the aperture bias. There is clear disagreement between these values and the mean values measured from real spectra.

Summarizing, both the evolution of the number density and of the stellar population ages of low-$\Sigma$ MPGs  strongly support a picture in which younger low-$\Sigma$ MPGs continuously appear at lower redshift. These results indicate that the increase both in number and in mean size of the population of MPGs is due to the continuous addition of larger and younger quiescent galaxies over cosmic time. 

\subsection{Comparison with previous work}

Evidence for older ages of the densest galaxies has been found by many authors both in the local Universe and at high-z \citep[e.g.][]{shankar09, saracco09, valentinuzzi10, williams10, saracco11, poggianti13a, carollo13, fagioli16}. However most of these works have investigated a larger stellar mass range, not focusing their analysis on the massive end.

In the same mass range, using UV colour to date stellar population ages of passive compact massive ellipticals, \citet{carollo13} found that they are consistent with passive evolution, in agreement with our results. Using a set of Lick absorption indices, \citet{onodera15} investigated stellar population properties for a sample of massive quiescent galaxies (UVJ colour selected) at $\langle z \rangle$ = 1.6.  They found a mean age of 1.1$^{+0.3}_{-0.2}$\,Gyr. As stated by the authors, this value is in excellent agreement with the age of local counterparts, if high-z massive quiescent galaxies evolve passively. We verified that more than 80$\%$ of their sample is composed of massive passive galaxies with $\Sigma >$ 2000\,M$_{\sun}$pc$^{-2}$, thus we can reasonably compare their results with ours. If we assume passive evolution between $z = 1.6$ and $z = 0.95$, the mean age of their massive quiescent galaxies rises to 3.1$^{+0.3}_{-0.2}$\,Gyr. This is in fair agreement with the mean age of high-$\Sigma$ MPGs we find at $z = 0.95$ (3.7$\pm$0.2\,Gyr) \citep[see also][]{whitaker13}.

Different conclusions were reached by F16. Using stacked spectra, the authors constrained the stellar population ages for dense and less dense passive galaxies in the zCOSMOS sample from $z = 0.8$ to $z = 0.2$. They found that the age of dense massive quiescent galaxies increases less than would be expected from passive evolution alone. Moreover, they found no correlation between the age of stellar population and the dimension of the source. The present analysis differs from the F16 analysis in a number of ways: the selection of passive galaxies (no emission lines+no MIPS vs NUV$rK$ colour); the selection of the dense sub-population (cut at fixed $\it{R_e}$ or along the SMR vs. cut at fixed $\Sigma$); and the procedure adopted to constrain the stellar population age. Unfortunately we cannot exactly reproduce their analysis with our data set. In Appendix B, we checked how our results change if we adopt the same criteria to select dense/less dense galaxies. We found that the mean age of dense MPGs is consistent with passive evolution and that less dense MPGs are younger than dense MPGs, independently of the criteria used to divide the sub-populations. However, as stated above, we stress that the previous checks rely on a sample of MPGs selected in a different way and adopt different techniques to constrain the stellar population age. In fact, we cannot repeat the same analysis with our data set, so we cannot fully account for the effect of these two factors on the results.

\section{Where do new large MPGs come from?}

\begin{figure}
 \begin{center}
  	\includegraphics[angle=0,width=9.0cm]{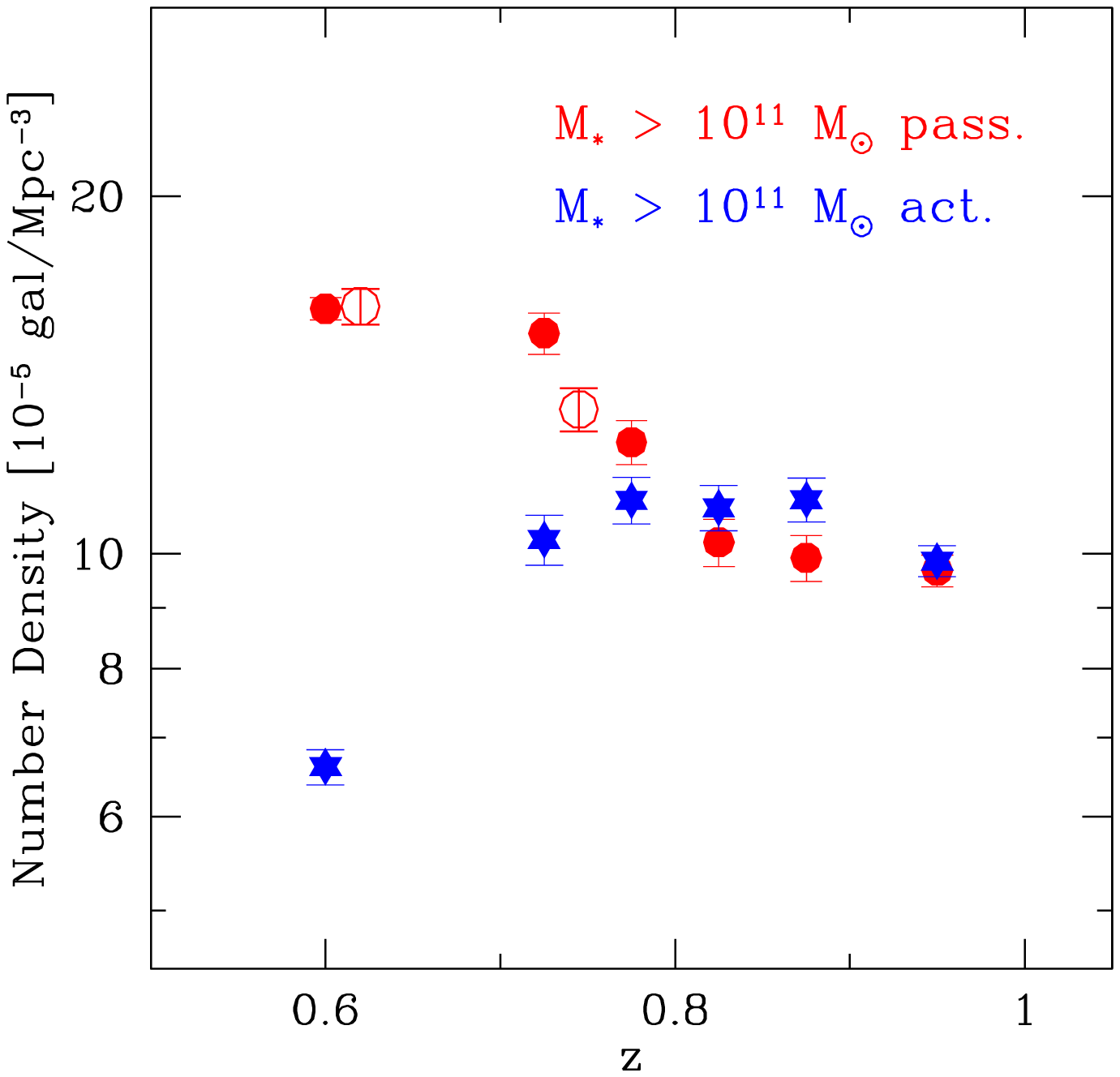} \\
\caption{The evolution of the number density of MPGs (filled red circles) and of star forming massive galaxies (MSFGs, blue filled stars). Open circles  show the expected growth in the abundance of MPGs below $z <$ 0.8, assuming that this is fully due to the observed decline of MSFGs. Solid and open circles have been shifted for visualisation purposes.}
	\label{actpass}
 \end{center}
\end{figure}

\begin{figure*}
 \begin{center}
  	\includegraphics[angle=90,width=18.0cm]{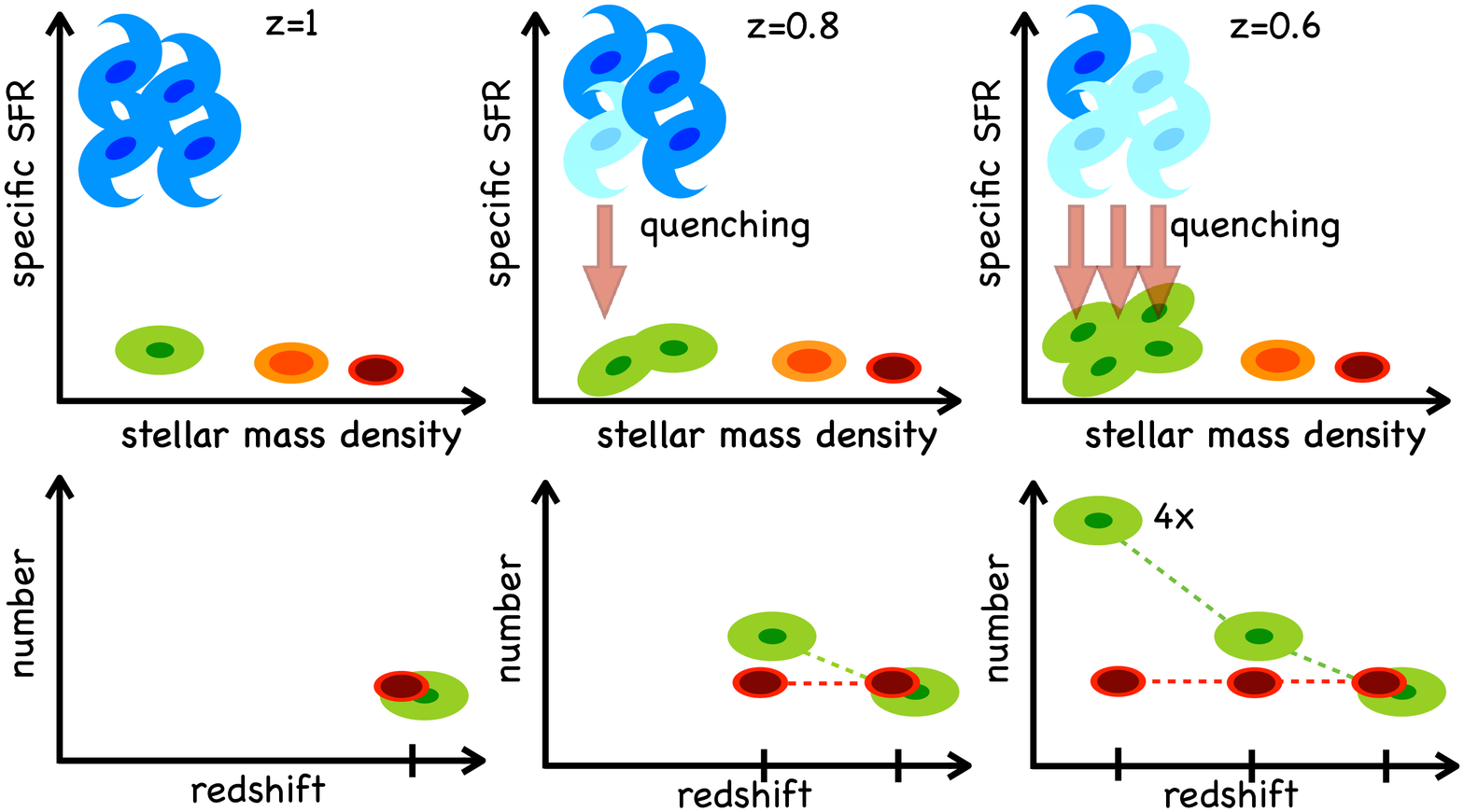} \\
	\caption{A schematic view of our results and findings. In the lower panels we plot the evolution of number density for low- and high-$\Sigma$ MPGs (large green and small red dark spheroids, respectively). From $z = 1.0$ to $z = 0.5$ the number density of high-$\Sigma$ MPGs does not increase, while over the same time period, the number of low-$\Sigma$ MPGs increases by a factor 4. These new low-$\Sigma$ MPGs are plausibly the direct descendants of MSFGs (blue spirals) that progressively halt their star formation until they become passive (see upper panels). In fact, we found that the observed increase in the number density of MPGs is totally accounted for by the observed decrease in the number density of MSFGs.}
	\label{cartoon}
 \end{center}
\end{figure*}

The evidence that the new MPGs are systematically younger contrasts with the hypothesis that massive galaxies are assembled mostly through dry mergers \citep[e.g.][]{hopkins09,cappellari12}. In fact, dry mergers should dilute any trend between stellar population age and time of appearance by mixing up the stellar population of pre-existing systems.  Actually, more recently new evidence has come to light supporting a scenario in which PGs are mostly the final evolutionary stage of star-forming galaxies (SFGs) that progressively halt their star formation until they become quiescent \citep[e.g.][]{lilly16, driver13}. In fact, at any $z$ and at fixed stellar mass, star-forming galaxies are larger than passive ones \citep[e.g.][]{vanderwel14}. Thus, if passive galaxies are just the quenched counterpart of star-forming galaxies, we should expect a correlation between the stellar population age of PGs and their mean stellar mass density in the direction of younger age for less dense systems. This is exactly what we have shown in Fig. \ref{d4evo}. In this case, however, further evidence is expected: the number of PGs that appear at any time, has to be similar to the number of SFGs that disappear. To test this simple hypothesis, in Fig. \ref{actpass} we compare the number density of the whole population of MPGs (red filled points), with the number density of massive star-forming galaxies (MSFGs) (blue filled stars). We considered all non passive massive galaxies to be MSFGs. The figure shows that the number densities of the two populations start to deviate at $z < 0.8$, i.e. when the Universe becomes efficient at producing low-$\Sigma$ MPGs.
In particular, the number density of MSFGs is almost flat at z$>$ 0.8 and then drops by a factor $\sim$2 at lower $z$. The declining trend does not depend on the selection criterion used to identify MSFGs. \citet{haines16} found the same trend studying  the number density of $\cal{M}\geqslant$ 10$^{11}$\,M$_{\odot}$ VIPERS galaxies with D4000$_{n} <$ 1.55. Starting from the very simple assumption that all the MSFGs that disappear at $z <$ 0.8, must necessarily migrate in the population of MPGs, we derived the number density of MPGs expected at $z < 0.8$ by considering their density at $z = 0.8$ and the observed decrement of MSFGs at $z <$ 0.8. Open red symbols report the results. In fact, they are in excellent agreement with the observed number density of MPGs. In this basic test we did not take into account the fact that some star-forming galaxies with $\cal{M} <$ 10$^{11}$\,M$_{\odot}$ can enter into our sample at a later time. Nevertheless, the comparison shown in Fig. \ref{actpass} shows that, to first order, the  migration of MSFGs to MPGs fully accounts for the increase in the number density of MPGs with cosmic time. In particular the $\langle$\,$\it{R_e}$\,$\rangle$ of MSFGs at z $\sim$ 0.8 is $\sim$ 5.7kpc, in agreement with the value $\langle$\,$\it{R_e}$\,$\rangle$ = 6.3$\pm$3kpc of low- and intermediate-$\Sigma$ MPGs at $\langle\,z\,\rangle$ = 0.5.
If a portion of low-$\Sigma$ MPGs assembled its stellar mass inside-out, i.e. starting from a compact passive core, we should add the number of these size-evolved MPGs to the open circles in Fig. \ref{actpass}, since the "progenitor" compact core is not included  in the population of MSFGs. In fact, including other channels for MPG production in the model would overestimate the number density of the population. We can therefore exclude the possibility that the inside-out accretion scenario is the main channel for the build up of MPG population. 
This evidence confirms in an independent way the results we found in Fig. \ref{numdentot} and Fig. \ref{d4evo} and that we summarize in the cartoon of Fig. \ref{cartoon}. 
In the $\sim$ 2.5\,Gyr of time from $z = 1.0$ to $z = 0.5$, the number density of high-$\Sigma$ MPGs does not evolve, and the ages of their stellar populations are consistent with a passive evolution. In the same time, the number density of low-$\Sigma$ MPGs increases by a factor 4. This increase is in fully consistent with the decrement we observe in the number density of MSFGs. If the new low-$\Sigma$ MPGs are the direct descendants of MSFGs, we should expect to find that the new low-$\Sigma$ MPGs would have to be younger than low-$\Sigma$ MPGs and high-$\Sigma$ MPGs already in place, that is what we found in Fig. \ref{d4evo}.

\section{Summary and Conclusions}

We used the VIPERS data set to investigate how the population of massive passive galaxies (MPGs) has been built up over cosmic time.

We looked at the evolution of both the number density and the mean age of the stellar population of MPGs as a function of redshift, and of the mean surface stellar mass density. From the VIPERS data set, we selected a sample of $\sim$ 2000 MPGs over the redshift range 0.5 $\leqslant z \leqslant$ 1.0. We divided this sample into three sub-populations according to their value of $\Sigma$:  high-$\Sigma$ MPGs, intermediate-$\Sigma$ MPGs, and low-$\Sigma$ MPGs. 

We studied the evolution of the number density for the three sub-populations of MPGs and found that it depends on $\Sigma$: the lower $\Sigma$ the faster the evolution (see Fig. \ref{numdentot}). In particular, we found that the number of dense galaxies per unit volume does not increase from $z = 1.0$ to $z = 0.5$. In fact, over the same time interval, the number density of less dense MPGs increases by a factor 4. This different evolution changes the composition of the population of MPGs with time. At $z >$ 0.8, high-, intermediate- and low-$\Sigma$ MPGs contribute equally to the population. At $z <$ 0.8, the number density of low-$\Sigma$ MPGs progressively increases and this sub-population starts to dominate over the other two classes. 

We then investigated the evolution of the stellar population ages as a function of $\Sigma$ (see Fig. \ref{d4evo}). We constrained the ages both using photometry, i.e. fitting the spectral energy distribution, and spectroscopy, through the D4000$_{n}$ index \citep{balogh99}. The two independent estimates are in agreement and show that:
\begin{itemize}
\item{The evolution of the age of high-$\Sigma$ MPGs is fully consistent with a passive ageing of their stellar population.}
\item{The evolution of the age of low-$\Sigma$ MPGs is slower than would be expected in the case of passive evolution, i.e. new low-$\Sigma$ MPGs are younger than existing ones.} 
\item{At any redshift, dense MPGs are older than less dense MPGs.}
\end{itemize}

Both the evolution of the number density and of the age of the stellar population of high-$\Sigma$ MPGs are consistent with passive evolution for this sub-population. On the other hand, the results we found for low-$\Sigma$ MPGs show that their number density continuously increases with decreasing redshift, i.e. new low-$\Sigma$ galaxies join the population of MPGs as time goes by. The study of stellar population age shows that these new galaxies are systematically younger than the low-$\Sigma$ MPGs already in place.

These results indicate that the increase both in number density and in typical radial size  observed for the population of MPGs is mostly due to the addition of less dense and younger galaxies at later times. Taking the results for low- and high-$\Sigma$ MPGs together we found that the population of MPGs was built by the continuous addition of less dense MPGs: on top of passively evolving dense MPGs already in place at $z = 1.0$, new, larger and younger quiescent galaxies continuously join the population of MPGs at later times. 

We find evidence that these new MPGs are the direct descendants of massive star-forming galaxies (MSFGs) that quenched their star formation. In fact, in Fig. \ref{actpass} we show that the observed increase in the number density of MPGs is totally accounted for by the decrease in the number density of MSFGs. This not only provides constraints on the origin of MPGs, but also rules out inside-out accretion as the main channel for their build up, confirming in an independent way our conclusions based on the evolution of number density and stellar population age.

\begin{acknowledgements}
We acknowledge the crucial contribution of the ESO staff for the management of service observations. In particular, we are deeply grateful to M. Hilker for his constant help and support of this program. Italian participation to VIPERS has been funded by INAF through PRIN 2008, 2010, and 2014 programs. LG, AJH, and BRG acknowledge support from the European Research Council through grant n.~291521. OLF acknowledges support from the European Research Council through grant n.~268107. TM and SA  acknowledge financial support from the ANR Spin(e) through the french grant  ANR-13-BS05-0005. AP, KM, and JK have been supported by the National Science Centre (grants UMO-2012/07/B/ST9/04425 and UMO-2013/09/D/ST9/04030). WJP is also grateful for support from the UK Science and Technology Facilities Council through the grant ST/I001204/1. EB, FM and LM acknowledge the support from grants ASI-INAF I/023/12/0 and PRIN MIUR 2010-2011. LM also acknowledges financial support from PRIN INAF 2012. SDLT acknowledges the support of the OCEVU Labex (ANR-11-LABX-0060) and the A*MIDEX project (ANR-11-IDEX-0001-02) funded by the "Investissements d'Avenir" French government program managed by the ANR. and the Programme National Galaxies et Cosmologie (PNCG). Research conducted within the scope of the HECOLS International Associated Laboratory, supported in part by the Polish NCN grant DEC-2013/08/M/ST9/00664.
\end{acknowledgements}

\nocite{}
\bibliographystyle{aa}
\bibliography{paper_I}

\begin{thebibliography}{81}
\expandafter\ifx\csname natexlab\endcsname\relax\def\natexlab#1{#1}\fi

\bibitem[{{Arnouts} {et~al.}(2013){Arnouts}, {Le Floc'h}, {Chevallard},
  {Johnson}, {Ilbert}, {Treyer}, {Aussel}, {Capak}, {Sanders}, {Scoville},
  {McCracken}, {Milliard}, {Pozzetti}, \& {Salvato}}]{arnouts13}
{Arnouts}, S., {Le Floc'h}, E., {Chevallard}, J., {et~al.} 2013, A$\&$A, 558,
  A67

\bibitem[{{Balogh} {et~al.}(1999){Balogh}, {Morris}, {Yee}, {Carlberg}, \&
  {Ellingson}}]{balogh99}
{Balogh}, M.~L., {Morris}, S.~L., {Yee}, H.~K.~C., {Carlberg}, R.~G., \&
  {Ellingson}, E. 1999, ApJ, 527, 54

\bibitem[{{Belli} {et~al.}(2014){Belli}, {Newman}, \& {Ellis}}]{belli14}
{Belli}, S., {Newman}, A.~B., \& {Ellis}, R.~S. 2014, ApJ, 783, 117

\bibitem[{{Bezanson} {et~al.}(2009){Bezanson}, {van Dokkum}, {Tal},
  {Marchesini}, {Kriek}, {Franx}, \& {Coppi}}]{bezanson09}
{Bezanson}, R., {van Dokkum}, P.~G., {Tal}, T., {et~al.} 2009, ApJ, 697, 1290

\bibitem[{{Bolzonella} {et~al.}(2000){Bolzonella}, {Miralles}, \&
  {Pell{\'o}}}]{bolzonella00}
{Bolzonella}, M., {Miralles}, J.-M., \& {Pell{\'o}}, R. 2000, A$\&$A, 363, 476

\bibitem[{{Brammer} {et~al.}(2011){Brammer}, {Whitaker}, {van Dokkum},
  {Marchesini}, {Franx}, {Kriek}, {Labb{\'e}}, {Lee}, {Muzzin}, {Quadri}, \&
  coauthors}]{brammer11}
{Brammer}, G.~B., {Whitaker}, K.~E., {van Dokkum}, P.~G., {et~al.} 2011, ApJ,
  739, 24

\bibitem[{{Bruzual} \& {Charlot}(2003)}]{bruzual03}
{Bruzual}, G. \& {Charlot}, S. 2003, MNRAS, 344, 1000

\bibitem[{{Calzetti} {et~al.}(2000){Calzetti}, {Armus}, {Bohlin}, {Kinney},
  {Koornneef}, \& {Storchi-Bergmann}}]{calzetti00}
{Calzetti}, D., {Armus}, L., {Bohlin}, R.~C., {et~al.} 2000, ApJ, 533, 682

\bibitem[{{Cappellari} {et~al.}(2011){Cappellari}, {Emsellem}, {Krajnovi{\'c}},
  {McDermid}, {Scott}, {Verdoes Kleijn}, {Young}, {Alatalo}, {Bacon}, {Blitz},
  \& coauthors}]{cappellari11}
{Cappellari}, M., {Emsellem}, E., {Krajnovi{\'c}}, D., {et~al.} 2011, MNRAS,
  413, 813

\bibitem[{{Cappellari} {et~al.}(2012){Cappellari}, {McDermid}, {Alatalo},
  {Blitz}, {Bois}, {Bournaud}, {Bureau}, {Crocker}, {Davies}, {Davis}, {de
  Zeeuw}, {Duc}, {Emsellem}, \& coauthors}]{cappellari12}
{Cappellari}, M., {McDermid}, R.~M., {Alatalo}, K., {et~al.} 2012, Nature, 484,
  485

\bibitem[{{Cappellari} {et~al.}(2013){Cappellari}, {Scott}, {Alatalo}, {Blitz},
  {Bois}, {Bournaud}, {Bureau}, {Crocker}, {Davies}, {Davis}, \&
  coauthors}]{cappellarixv}
{Cappellari}, M., {Scott}, N., {Alatalo}, K., {et~al.} 2013, MNRAS, 432, 1709

\bibitem[{{Carollo} {et~al.}(2013){Carollo}, {Bschorr}, {Renzini}, {Lilly},
  {Capak}, {Cibinel}, {Ilbert}, {Onodera}, {Scoville}, {Cameron}, {Mobasher},
  {Sanders}, \& {Taniguchi}}]{carollo13}
{Carollo}, C.~M., {Bschorr}, T.~J., {Renzini}, A., {et~al.} 2013, ApJ, 773, 112

\bibitem[{{Cassata} {et~al.}(2010){Cassata}, {Giavalisco}, {Guo}, {Ferguson},
  {Koekemoer}, {Renzini}, {Fontana}, {Salimbeni}, {Dickinson}, {Casertano},
  {Conselice}, {Grogin}, {Lotz}, {Papovich}, {Lucas}, {Straughn}, {Gardner}, \&
  {Moustakas}}]{cassata10}
{Cassata}, P., {Giavalisco}, M., {Guo}, Y., {et~al.} 2010, ApJ, 714, L79

\bibitem[{{Cassata} {et~al.}(2011){Cassata}, {Giavalisco}, {Guo}, {Renzini},
  {Ferguson}, {Koekemoer}, {Salimbeni}, {Scarlata}, {Grogin}, {Conselice},
  {Dahlen}, {Lotz}, {Dickinson}, \& {Lin}}]{cassata11}
{Cassata}, P., {Giavalisco}, M., {Guo}, Y., {et~al.} 2011, ApJ, 743, 96

\bibitem[{{Cassata} {et~al.}(2013){Cassata}, {Giavalisco}, {Williams}, {Guo},
  {Lee}, {Renzini}, {Ferguson}, {Faber}, {Barro}, {McIntosh}, \&
  coauthors}]{cassata13}
{Cassata}, P., {Giavalisco}, M., {Williams}, C.~C., {et~al.} 2013, ApJ, 775,
  106

\bibitem[{{Chabrier}(2003)}]{chabrier03}
{Chabrier}, G. 2003, PASP, 115, 763

\bibitem[{{Cimatti} {et~al.}(2008){Cimatti}, {Cassata}, {Pozzetti}, {Kurk},
  {Mignoli}, {Renzini}, {Daddi}, {Bolzonella}, {Brusa}, {Rodighiero},
  {Dickinson}, {Franceschini}, {Zamorani}, {Berta}, {Rosati}, \&
  {Halliday}}]{cimatti08}
{Cimatti}, A., {Cassata}, P., {Pozzetti}, L., {et~al.} 2008, A$\&$A, 482, 21

\bibitem[{{Daddi} {et~al.}(2005){Daddi}, {Renzini}, {Pirzkal}, {Cimatti},
  {Malhotra}, {Stiavelli}, {Xu}, {Pasquali}, {Rhoads}, {Brusa}, {di Serego
  Alighieri}, {Ferguson}, {Koekemoer}, {Moustakas}, {Panagia}, \&
  {Windhorst}}]{daddi05}
{Daddi}, E., {Renzini}, A., {Pirzkal}, N., {et~al.} 2005, ApJ, 626, 680

\bibitem[{{Damjanov} {et~al.}(2011){Damjanov}, {Abraham}, {Glazebrook},
  {McCarthy}, {Caris}, {Carlberg}, {Chen}, {Crampton}, {Green}, {J{\o}rgensen},
  {Juneau}, {Le Borgne}, {Marzke}, {Mentuch}, {Murowinski}, {Roth}, {Savaglio},
  \& {Yan}}]{damjanov11}
{Damjanov}, I., {Abraham}, R.~G., {Glazebrook}, K., {et~al.} 2011, ApJL, 739,
  L44

\bibitem[{{Damjanov} {et~al.}(2015){Damjanov}, {Geller}, {Zahid}, \&
  {Hwang}}]{damjanov15}
{Damjanov}, I., {Geller}, M.~J., {Zahid}, H.~J., \& {Hwang}, H.~S. 2015, \apj,
  806, 158

\bibitem[{{Damjanov} {et~al.}(2014){Damjanov}, {Hwang}, {Geller}, \&
  {Chilingarian}}]{damjanov14}
{Damjanov}, I., {Hwang}, H.~S., {Geller}, M.~J., \& {Chilingarian}, I. 2014,
  ApJ, 793, 39

\bibitem[{{Davidzon} {et~al.}(2013){Davidzon}, {Bolzonella}, {Coupon},
  {Ilbert}, {Arnouts}, {de la Torre}, {Fritz}, {De Lucia}, {Iovino}, {Granett},
  \& coauthors}]{davidzon13}
{Davidzon}, I., {Bolzonella}, M., {Coupon}, J., {et~al.} 2013, A$\&$A, 558, A23

\bibitem[{{Davidzon} {et~al.}(2016){Davidzon}, {Cucciati}, {Bolzonella}, {De
  Lucia}, {Zamorani}, {Arnouts}, {Moutard}, {Ilbert}, {Garilli}, {Scodeggio},
  {Guzzo}, {Abbas}, {Adami}, {Bel}, {Bottini}, {Branchini}, {Cappi}, {Coupon},
  {de la Torre}, {Di Porto}, {Fritz}, {Franzetti}, {Fumana}, {Granett},
  {Guennou}, {Iovino}, {Krywult}, {Le Brun}, {Le F{\`e}vre}, {Maccagni},
  {Ma{\l}ek}, {Marulli}, {McCracken}, {Mellier}, {Moscardini}, {Polletta},
  {Pollo}, {Tasca}, {Tojeiro}, {Vergani}, \& {Zanichelli}}]{davidzon16}
{Davidzon}, I., {Cucciati}, O., {Bolzonella}, M., {et~al.} 2016, A$\&$A, 586,
  A23

\bibitem[{{De Lucia} \& {Blaizot}(2007)}]{delucia07}
{De Lucia}, G. \& {Blaizot}, J. 2007, MNRAS, 375, 2

\bibitem[{{Dekel} {et~al.}(2009){Dekel}, {Birnboim}, {Engel}, {Freundlich},
  {Goerdt}, {Mumcuoglu}, {Neistein}, {Pichon}, {Teyssier}, \&
  {Zinger}}]{dekel09}
{Dekel}, A., {Birnboim}, Y., {Engel}, G., {et~al.} 2009, Nature, 457, 451

\bibitem[{{Driver} {et~al.}(2013){Driver}, {Robotham}, {Bland-Hawthorn},
  {Brown}, {Hopkins}, {Liske}, {Phillipps}, \& {Wilkins}}]{driver13}
{Driver}, S.~P., {Robotham}, A.~S.~G., {Bland-Hawthorn}, J., {et~al.} 2013,
  MNRAS, 430, 2622

\bibitem[{{Fagioli} {et~al.}(2016){Fagioli}, {Carollo}, {Renzini}, {Lilly},
  {Onodera}, \& {Tacchella}}]{fagioli16}
{Fagioli}, M., {Carollo}, C.~M., {Renzini}, A., {et~al.} 2016, \apj, 831, 173

\bibitem[{{Franx} \& {van Dokkum}(1996)}]{franx96}
{Franx}, M. \& {van Dokkum}, P.~G. 1996, in IAU Symposium, Vol. 171, New Light
  on Galaxy Evolution, ed. R.~{Bender} \& R.~L. {Davies}, 233

\bibitem[{{Fritz} {et~al.}(2014){Fritz}, {Scodeggio}, {Ilbert}, {Bolzonella},
  {Davidzon}, {Coupon}, {Garilli}, {Guzzo}, {Zamorani}, {Abbas}, \&
  coauthors}]{fritz14}
{Fritz}, A., {Scodeggio}, M., {Ilbert}, O., {et~al.} 2014, A$\&$A, 563, A92

\bibitem[{{Gallazzi} {et~al.}(2014){Gallazzi}, {Bell}, {Zibetti}, {Brinchmann},
  \& {Kelson}}]{gallazzi14}
{Gallazzi}, A., {Bell}, E.~F., {Zibetti}, S., {Brinchmann}, J., \& {Kelson},
  D.~D. 2014, ApJ, 788, 72

\bibitem[{{Gallazzi} {et~al.}(2005){Gallazzi}, {Charlot}, {Brinchmann},
  {White}, \& {Tremonti}}]{gallazzi05}
{Gallazzi}, A., {Charlot}, S., {Brinchmann}, J., {White}, S.~D.~M., \&
  {Tremonti}, C.~A. 2005, MNRAS, 362, 41

\bibitem[{{Gargiulo} {et~al.}(2012){Gargiulo}, {Saracco}, {Longhetti}, {La
  Barbera}, \& {Tamburri}}]{gargiulo12}
{Gargiulo}, A., {Saracco}, P., {Longhetti}, M., {La Barbera}, F., \&
  {Tamburri}, S. 2012, MNRAS, 425, 2698

\bibitem[{{Gargiulo} {et~al.}(2016){Gargiulo}, {Saracco}, {Tamburri}, {Lonoce},
  \& {Ciocca}}]{gargiulo16}
{Gargiulo}, A., {Saracco}, P., {Tamburri}, S., {Lonoce}, I., \& {Ciocca}, F.
  2016, A$\&$A, 592, A132

\bibitem[{{Garilli} {et~al.}(2014){Garilli}, {Guzzo}, {Scodeggio},
  {Bolzonella}, {Abbas}, {Adami}, {Arnouts}, {Bel}, {Bottini}, {Branchini}, \&
  coauthors}]{garilli14}
{Garilli}, B., {Guzzo}, L., {Scodeggio}, M., {et~al.} 2014, A$\&$A, 562, A23

\bibitem[{{Guo} {et~al.}(2011){Guo}, {Giavalisco}, {Cassata}, {Ferguson},
  {Dickinson}, {Renzini}, {Koekemoer}, {Grogin}, {Papovich}, {Tundo},
  {Fontana}, {Lotz}, \& {Salimbeni}}]{guo11}
{Guo}, Y., {Giavalisco}, M., {Cassata}, P., {et~al.} 2011, ApJ, 735, 18

\bibitem[{{Guzzo} {et~al.}(2014){Guzzo}, {Scodeggio}, {Garilli}, {Granett},
  {Fritz}, {Abbas}, {Adami}, {Arnouts}, {Bel}, {Bolzonella}, \&
  authors}]{guzzo14}
{Guzzo}, L., {Scodeggio}, M., {Garilli}, B., {et~al.} 2014, A$\&$A, 566, A108

\bibitem[{{Haines} {et~al.}(2016)}]{haines16}
{Haines}, C. {et~al.} 2016, \aap~\rm{submitted}

\bibitem[{{Hilz} {et~al.}(2013){Hilz}, {Naab}, \& {Ostriker}}]{hilz13}
{Hilz}, M., {Naab}, T., \& {Ostriker}, J.~P. 2013, MNRAS, 429, 2924

\bibitem[{{Hopkins} {et~al.}(2009){Hopkins}, {Bundy}, {Murray}, {Quataert},
  {Lauer}, \& {Ma}}]{hopkins09}
{Hopkins}, P.~F., {Bundy}, K., {Murray}, N., {et~al.} 2009, MNRAS, 398, 898

\bibitem[{{Hopkins} {et~al.}(2008){Hopkins}, {Cox}, \& {Hernquist}}]{hopkins08}
{Hopkins}, P.~F., {Cox}, T.~J., \& {Hernquist}, L. 2008, ApJ, 689, 17

\bibitem[{{Ilbert} {et~al.}(2010){Ilbert}, {Salvato}, {Le Floc'h}, {Aussel},
  {Capak}, {McCracken}, {Mobasher}, {Kartaltepe}, {Scoville}, {Sanders},
  {Arnouts}, {Bundy}, {Cassata}, {Kneib}, {Koekemoer}, {Le F{\`e}vre}, {Lilly},
  {Surace}, {Taniguchi}, {Tasca}, {Thompson}, {Tresse}, {Zamojski}, {Zamorani},
  \& {Zucca}}]{ilbert10}
{Ilbert}, O., {Salvato}, M., {Le Floc'h}, E., {et~al.} 2010, ApJ, 709, 644

\bibitem[{{Jarvis} {et~al.}(2013){Jarvis}, {Bonfield}, {Bruce}, {Geach},
  {McAlpine}, {McLure}, {Gonz{\'a}lez-Solares}, {Irwin}, {Lewis}, {Yoldas}, \&
  coauthors}]{jarvis13}
{Jarvis}, M.~J., {Bonfield}, D.~G., {Bruce}, V.~A., {et~al.} 2013, \mnras, 428,
  1281

\bibitem[{{Kauffmann} {et~al.}(2003){Kauffmann}, {Heckman}, {White}, {Charlot},
  {Tremonti}, {Brinchmann}, {Bruzual}, {Peng}, {Seibert}, {Bernardi}, \&
  coauthors}]{kauffmann03}
{Kauffmann}, G., {Heckman}, T.~M., {White}, S.~D.~M., {et~al.} 2003, MNRAS,
  341, 33

\bibitem[{{Kere{\v s}} {et~al.}(2005){Kere{\v s}}, {Katz}, {Weinberg}, \&
  {Dav{\'e}}}]{keres05}
{Kere{\v s}}, D., {Katz}, N., {Weinberg}, D.~H., \& {Dav{\'e}}, R. 2005, MNRAS,
  363, 2

\bibitem[{{Krywult} {et~al.}(2016){Krywult}, {Tasca}, {Pollo}, {Vergani},
  {Bolzonella}, {Davidzon}, {Iovino}, {Gargiulo}, {Haines}, {Scodeggio}, \&
  coauthors}]{krywult16}
{Krywult}, J., {Tasca}, L.~A.~M., {Pollo}, A., {et~al.} 2016, ArXiv e-prints
  [\eprint[arXiv]{1605.05502}]

\bibitem[{{La Barbera} \& {de Carvalho}(2009)}]{labarbera09}
{La Barbera}, F. \& {de Carvalho}, R.~R. 2009, ApJ, 699, L76

\bibitem[{{Lilly} \& {Carollo}(2016)}]{lilly16}
{Lilly}, S.~J. \& {Carollo}, C.~M. 2016, ArXiv e-prints
  [\eprint[arXiv]{1604.06459}]

\bibitem[{{Lilly} {et~al.}(2007){Lilly}, {Le F{\`e}vre}, {Renzini}, {Zamorani},
  {Scodeggio}, {Contini}, {Carollo}, {Hasinger}, {Kneib}, {Iovino}, \&
  coauthors}]{lilly07}
{Lilly}, S.~J., {Le F{\`e}vre}, O., {Renzini}, A., {et~al.} 2007, ApJS, 172, 70

\bibitem[{{Longhetti} {et~al.}(2007){Longhetti}, {Saracco}, {Severgnini},
  {Della Ceca}, {Mannucci}, {Bender}, {Drory}, {Feulner}, \&
  {Hopp}}]{longhetti07}
{Longhetti}, M., {Saracco}, P., {Severgnini}, P., {et~al.} 2007, MNRAS, 374,
  614

\bibitem[{{McDermid} {et~al.}(2015){McDermid}, {Alatalo}, {Blitz}, {Bournaud},
  {Bureau}, {Cappellari}, {Crocker}, {Davies}, {Davis}, {de Zeeuw}, \&
  coauthors}]{mcdermid15}
{McDermid}, R.~M., {Alatalo}, K., {Blitz}, L., {et~al.} 2015, MNRAS, 448, 3484

\bibitem[{{Moresco} {et~al.}(2013){Moresco}, {Pozzetti}, {Cimatti}, {Zamorani},
  {Bolzonella}, {Lamareille}, {Mignoli}, {Zucca}, {Lilly}, {Carollo}, \&
  coauthors}]{moresco13}
{Moresco}, M., {Pozzetti}, L., {Cimatti}, A., {et~al.} 2013, A$\&$A, 558, A61

\bibitem[{{Moutard} {et~al.}(2016{\natexlab{a}}){Moutard}, {Arnouts}, {Ilbert},
  {Coupon}, {Davidzon}, {Guzzo}, \& coauthors}]{moutard16}
{Moutard}, T., {Arnouts}, S., {Ilbert}, O., {et~al.} 2016{\natexlab{a}},
  A$\&$A, 590, A103

\bibitem[{{Moutard} {et~al.}(2016{\natexlab{b}}){Moutard}, {Arnouts}, {Ilbert},
  {Coupon}, {Davidzon}, {Guzzo}, {Hudelot}, {McCracken}, {Van Werbaeke},
  {Morrison}, \& coauthors}]{moutard16a}
{Moutard}, T., {Arnouts}, S., {Ilbert}, O., {et~al.} 2016{\natexlab{b}},
  A$\&$A, 590, A103

\bibitem[{{Naab} {et~al.}(2009){Naab}, {Johansson}, \& {Ostriker}}]{naab09}
{Naab}, T., {Johansson}, P.~H., \& {Ostriker}, J.~P. 2009, ApJ, 699, L178

\bibitem[{{Naab} {et~al.}(2007){Naab}, {Johansson}, {Ostriker}, \&
  {Efstathiou}}]{naab07}
{Naab}, T., {Johansson}, P.~H., {Ostriker}, J.~P., \& {Efstathiou}, G. 2007,
  ApJ, 658, 710

\bibitem[{{Onodera} {et~al.}(2015){Onodera}, {Carollo}, {Renzini},
  {Cappellari}, {Mancini}, {Arimoto}, {Daddi}, {Gobat}, {Strazzullo},
  {Tacchella}, \& {Yamada}}]{onodera15}
{Onodera}, M., {Carollo}, C.~M., {Renzini}, A., {et~al.} 2015, ApJ, 808, 161

\bibitem[{{Peng} {et~al.}(2002){Peng}, {Ho}, {Impey}, \& {Rix}}]{peng02}
{Peng}, C.~Y., {Ho}, L.~C., {Impey}, C.~D., \& {Rix}, H. 2002, AJ, 124, 266

\bibitem[{{Poggianti} {et~al.}(2013{\natexlab{a}}){Poggianti}, {Calvi},
  {Bindoni}, {D'Onofrio}, {Moretti}, {Valentinuzzi}, {Fasano}, {Fritz}, {De
  Lucia}, {Vulcani}, {Bettoni}, {Gullieuszik}, \& {Omizzolo}}]{poggianti13a}
{Poggianti}, B.~M., {Calvi}, R., {Bindoni}, D., {et~al.} 2013{\natexlab{a}},
  ApJ, 762, 77

\bibitem[{{Poggianti} {et~al.}(2013{\natexlab{b}}){Poggianti}, {Moretti},
  {Calvi}, {D'Onofrio}, {Valentinuzzi}, {Fritz}, \& {Renzini}}]{poggianti13}
{Poggianti}, B.~M., {Moretti}, A., {Calvi}, R., {et~al.} 2013{\natexlab{b}},
  ApJ, 777, 125

\bibitem[{{Pozzetti} {et~al.}(2010){Pozzetti}, {Bolzonella}, {Zucca},
  {Zamorani}, {Lilly}, {Renzini}, {Moresco}, {Mignoli}, {Cassata}, {Tasca}, \&
  coauthors}]{pozzetti10}
{Pozzetti}, L., {Bolzonella}, M., {Zucca}, E., {et~al.} 2010, A$\&$A, 523, A13

\bibitem[{{Prevot} {et~al.}(1984){Prevot}, {Lequeux}, {Prevot}, {Maurice}, \&
  {Rocca-Volmerange}}]{prevot84}
{Prevot}, M.~L., {Lequeux}, J., {Prevot}, L., {Maurice}, E., \&
  {Rocca-Volmerange}, B. 1984, A$\&$A, 132, 389

\bibitem[{{Saglia} {et~al.}(2010){Saglia}, {Fabricius}, {Bender}, {Montalto},
  {Lee}, {Riffeser}, {Seitz}, {Morganti}, {Gerhard}, \& {Hopp}}]{saglia10}
{Saglia}, R.~P., {Fabricius}, M., {Bender}, R., {et~al.} 2010, A$\&$A, 509, A61

\bibitem[{{Saracco} {et~al.}(2009){Saracco}, {Longhetti}, \&
  {Andreon}}]{saracco09}
{Saracco}, P., {Longhetti}, M., \& {Andreon}, S. 2009, MNRAS, 392, 718

\bibitem[{{Saracco} {et~al.}(2010){Saracco}, {Longhetti}, \&
  {Gargiulo}}]{saracco10}
{Saracco}, P., {Longhetti}, M., \& {Gargiulo}, A. 2010, MNRAS, L115+

\bibitem[{{Saracco} {et~al.}(2011){Saracco}, {Longhetti}, \&
  {Gargiulo}}]{saracco11}
{Saracco}, P., {Longhetti}, M., \& {Gargiulo}, A. 2011, MNRAS, 412, 2707

\bibitem[{{Scodeggio}(2016)}]{scodeggio16}
{Scodeggio}, M., e.~a. 2016, \aap~\rm{submitted}

\bibitem[{{Scoville} {et~al.}(2007){Scoville}, {Aussel}, {Brusa}, {Capak},
  {Carollo}, {Elvis}, {Giavalisco}, {Guzzo}, {Hasinger}, {Impey}, \&
  coauthors}]{scoville07}
{Scoville}, N., {Aussel}, H., {Brusa}, M., {et~al.} 2007, ApJS, 172, 1

\bibitem[{{Shankar} \& {Bernardi}(2009)}]{shankar09}
{Shankar}, F. \& {Bernardi}, M. 2009, MNRAS, 396, L76

\bibitem[{{Shen} {et~al.}(2003){Shen}, {Mo}, {White}, {Blanton}, {Kauffmann},
  {Voges}, {Brinkmann}, \& {Csabai}}]{shen03}
{Shen}, S., {Mo}, H.~J., {White}, S.~D.~M., {et~al.} 2003, MNRAS, 343, 978

\bibitem[{{Siudek} {et~al.}(2016){Siudek}, {Ma{\l}ek}, {Scodeggio}, {Garilli},
  {Pollo}, {Haines}, {Fritz}, {Bolzonella}, {de la Torre}, {Granett}, \&
  coauthors}]{siudek16}
{Siudek}, M., {Ma{\l}ek}, K., {Scodeggio}, M., {et~al.} 2016, ArXiv e-prints
  [\eprint[arXiv]{1605.05503}]

\bibitem[{{Spolaor} {et~al.}(2010){Spolaor}, {Kobayashi}, {Forbes}, {Couch}, \&
  {Hau}}]{spolaor10}
{Spolaor}, M., {Kobayashi}, C., {Forbes}, D.~A., {Couch}, W.~J., \& {Hau},
  G.~K.~T. 2010, MNRAS, 408, 272

\bibitem[{{Tamburri} {et~al.}(2014){Tamburri}, {Saracco}, {Longhetti},
  {Gargiulo}, {Lonoce}, \& {Ciocca}}]{tamburri14}
{Tamburri}, S., {Saracco}, P., {Longhetti}, M., {et~al.} 2014, \aap, 570, A102

\bibitem[{{Trujillo} {et~al.}(2006){Trujillo}, {Feulner}, {Goranova}, {Hopp},
  {Longhetti}, {Saracco}, {Bender}, {Braito}, {Della Ceca}, {Drory},
  {Mannucci}, \& {Severgnini}}]{trujillo06}
{Trujillo}, I., {Feulner}, G., {Goranova}, Y., {et~al.} 2006, MNRAS, 373, L36

\bibitem[{{Valentinuzzi} {et~al.}(2010){Valentinuzzi}, {Fritz}, {Poggianti},
  {Cava}, {Bettoni}, {Fasano}, {D'Onofrio}, {Couch}, {Dressler}, {Moles},
  {Moretti}, {Omizzolo}, {Kj{\ae}rgaard}, {Vanzella}, \&
  {Varela}}]{valentinuzzi10}
{Valentinuzzi}, T., {Fritz}, J., {Poggianti}, B.~M., {et~al.} 2010, ApJ, 712,
  226

\bibitem[{{van der Wel} {et~al.}(2014){van der Wel}, {Franx}, {van Dokkum},
  {Skelton}, {Momcheva}, {Whitaker}, {Brammer}, {Bell}, {Rix}, {Wuyts},
  {Ferguson}, {Holden}, {Barro}, {Koekemoer}, {Chang}, {McGrath},
  {H{\"a}ussler}, {Dekel}, {Behroozi}, {Fumagalli}, {Leja}, {Lundgren},
  {Maseda}, {Nelson}, {Wake}, {Patel}, {Labb{\'e}}, {Faber}, {Grogin}, \&
  {Kocevski}}]{vanderwel14}
{van der Wel}, A., {Franx}, M., {van Dokkum}, P.~G., {et~al.} 2014, \apj, 788,
  28

\bibitem[{{van der Wel} {et~al.}(2008){van der Wel}, {Holden}, {Zirm}, {Franx},
  {Rettura}, {Illingworth}, \& {Ford}}]{vanderwel08}
{van der Wel}, A., {Holden}, B.~P., {Zirm}, A.~W., {et~al.} 2008, ApJ, 688, 48

\bibitem[{{van Dokkum} {et~al.}(2008){van Dokkum}, {Franx}, {Kriek}, {Holden},
  {Illingworth}, {Magee}, {Bouwens}, {Marchesini}, {Quadri}, {Rudnick},
  {Taylor}, \& {Toft}}]{vandokkum08}
{van Dokkum}, P.~G., {Franx}, M., {Kriek}, M., {et~al.} 2008, ApJ, 677, L5

\bibitem[{{Whitaker} {et~al.}(2013){Whitaker}, {van Dokkum}, {Brammer},
  {Momcheva}, {Skelton}, {Franx}, {Kriek}, {Labb{\'e}}, {Fumagalli},
  {Lundgren}, {Nelson}, {Patel}, \& {Rix}}]{whitaker13}
{Whitaker}, K.~E., {van Dokkum}, P.~G., {Brammer}, G., {et~al.} 2013, ApJL,
  770, L39

\bibitem[{{Williams} {et~al.}(2009){Williams}, {Quadri}, {Franx}, {van Dokkum},
  \& {Labb{\'e}}}]{williams09}
{Williams}, R.~J., {Quadri}, R.~F., {Franx}, M., {van Dokkum}, P., \&
  {Labb{\'e}}, I. 2009, \apj, 691, 1879

\bibitem[{{Williams} {et~al.}(2010){Williams}, {Quadri}, {Franx}, {van Dokkum},
  {Toft}, {Kriek}, \& {Labb{\'e}}}]{williams10}
{Williams}, R.~J., {Quadri}, R.~F., {Franx}, M., {et~al.} 2010, ApJ, 713, 738

\bibitem[{{Zahid} {et~al.}(2016){Zahid}, {Baeza Hochmuth}, {Geller},
  {Damjanov}, {Chillingarian}, {Sohn}, {Salmi}, \& {Hwang}}]{zahid16}
{Zahid}, H.~J., {Baeza Hochmuth}, N., {Geller}, M.~J., {et~al.} 2016, ArXiv
  e-prints [\eprint[arXiv]{1605.09734}]

\end{thebibliography}

\begin{appendix}
 \appendix
 
\section{The dependence of D4000$_{n}$ on metallicity, timescale of star formation and resolution.}

\begin{figure}
 \begin{center}
\includegraphics[angle=0,width=8.5cm]{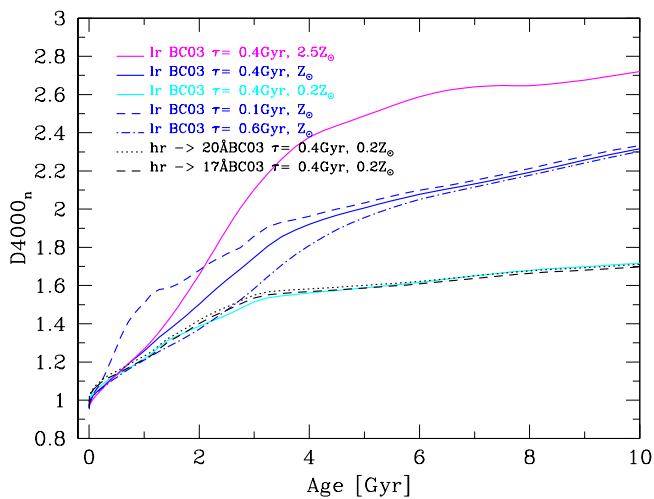}
	\caption{The D4000$_{n}$ index as a function of the age for BC03 low resolution models with exponentially declining star formation history and different $tau$ and Z. Magenta, blue and cyan lines correspond to models with $\tau$ = 0.4 Gyr and Z = 2Z$_{\sun}$, Z$_{\sun}$, and 0.2Z$_{\sun}$. Dashed and dot-dashed blue lines indicate models with Z = Z$_{\sun}$ and $tau$ = 0.1 Gyr and $tau$ = 0.6 Gyr, respectively. Dotted and dashed black lines indicate models with Z = 02Z$_{\sun}$, $\tau$ = 0.4 Gyr and resolution 20\smash{AA}\, and 17\smash{AA}\, respectively.
    }
	\label{models}
 \end{center}
\end{figure}

In Fig. \ref{models} we report D4000$_{n}$ for different BC03 models in order to show the dependence
of this quantity on metallicity, timescale of star formation and spectral resolution.
Cyan, blue and magenta solid lines track the evolution of D4000$_{n}$ with stellar population age  
for BC03 low resolution (lr) models with the same star formation timescale ($\tau$ = 0.4\,Gyr) and sub-solar (0.2Z$_{\odot}$) metallicity, 
solar metallicity, and super-solar (2.5Z$_{\odot}$) metallicity respectively.
Dashed and dot-dashed blue lines show the evolution of D4000$_{n}$ for models with solar metallicity and $\tau$ = 0.1\,Gyr and 
$\tau$ = 0.6\,Gyr, respectively. Finally, dotted and dashed black lines show the D4000$_{n}$ for two models with the same 
$\tau$  and metallicity but different resolution.
In particular, starting from a high resolution BC03 model with $\tau$ = 0.4\,Gyr and Z = 0.2Z$_{\odot}$ we downgrade it 
to 20\smash{\AA}\, (dotted line) and to 17\smash{\AA}\, (dashed line), the typical resolution of VIPERS spectra.

\section{Comparison with the analysis by Fagioli et al.}

\begin{figure}
 \begin{center}
 \begin{tabular}{cc}
\includegraphics[angle=0,width=4.1cm]{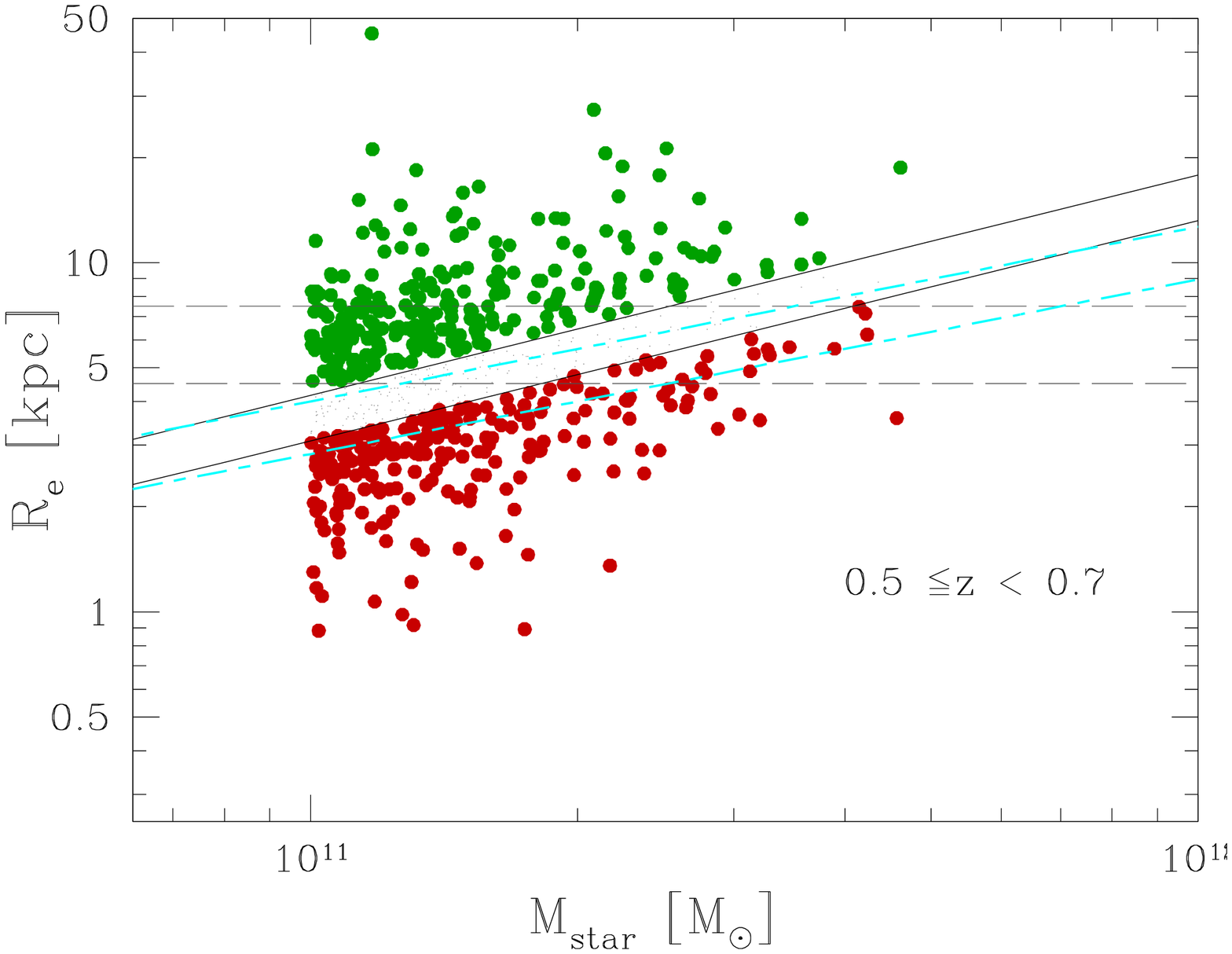} &  
\includegraphics[angle=0,width=4.1cm]{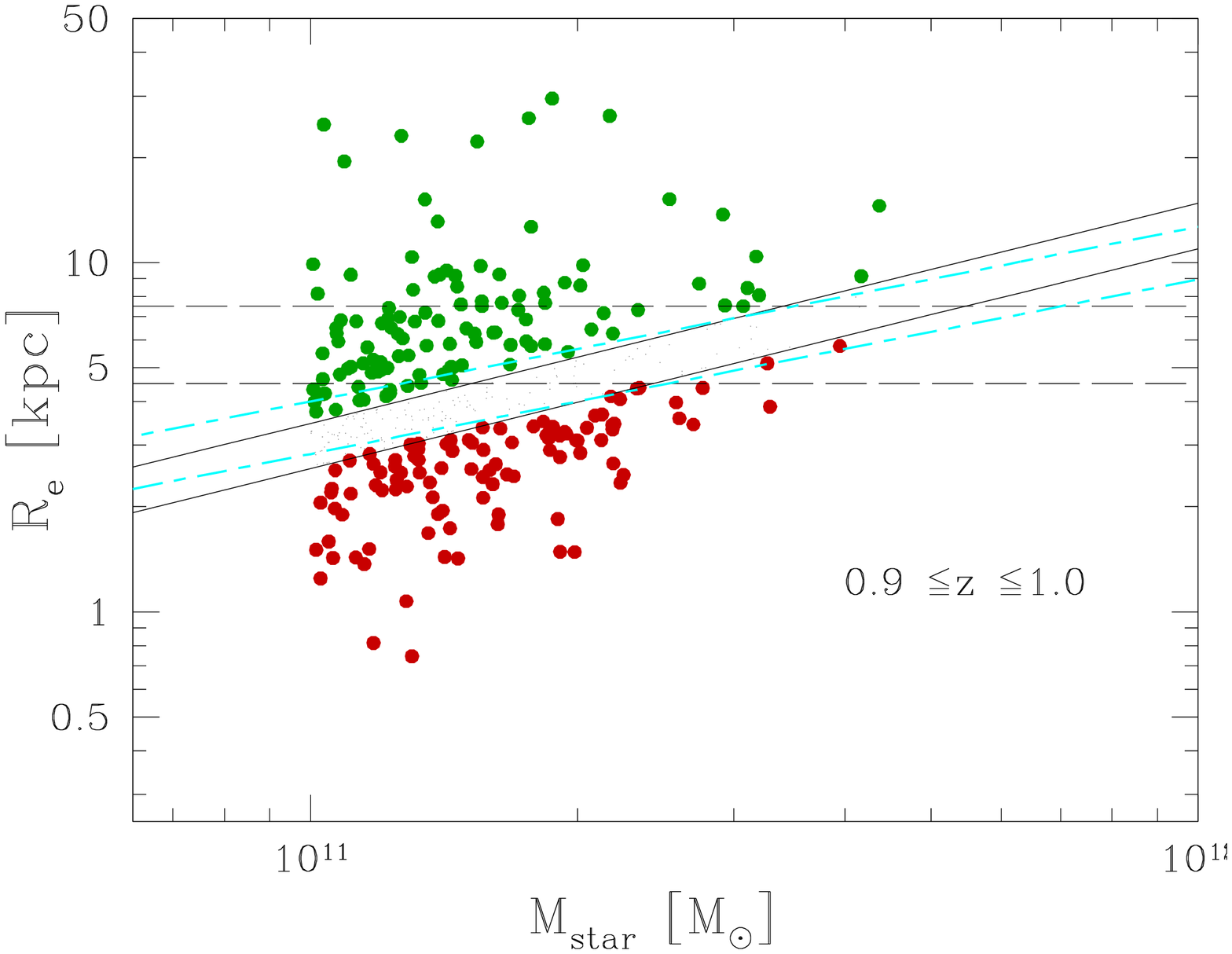}\\
\includegraphics[angle=0,width=4.1cm]{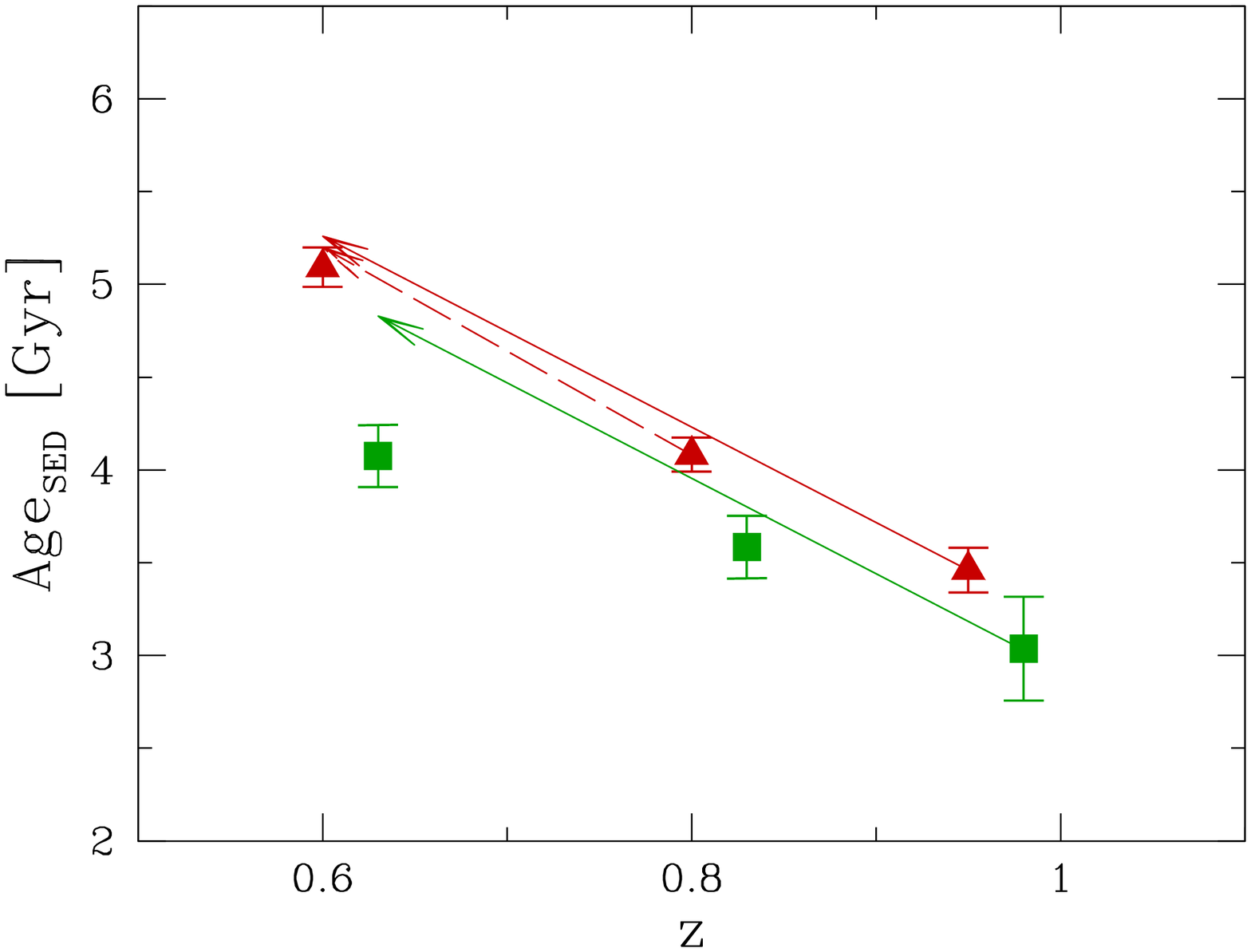} &  
\includegraphics[angle=0,width=4.1cm]{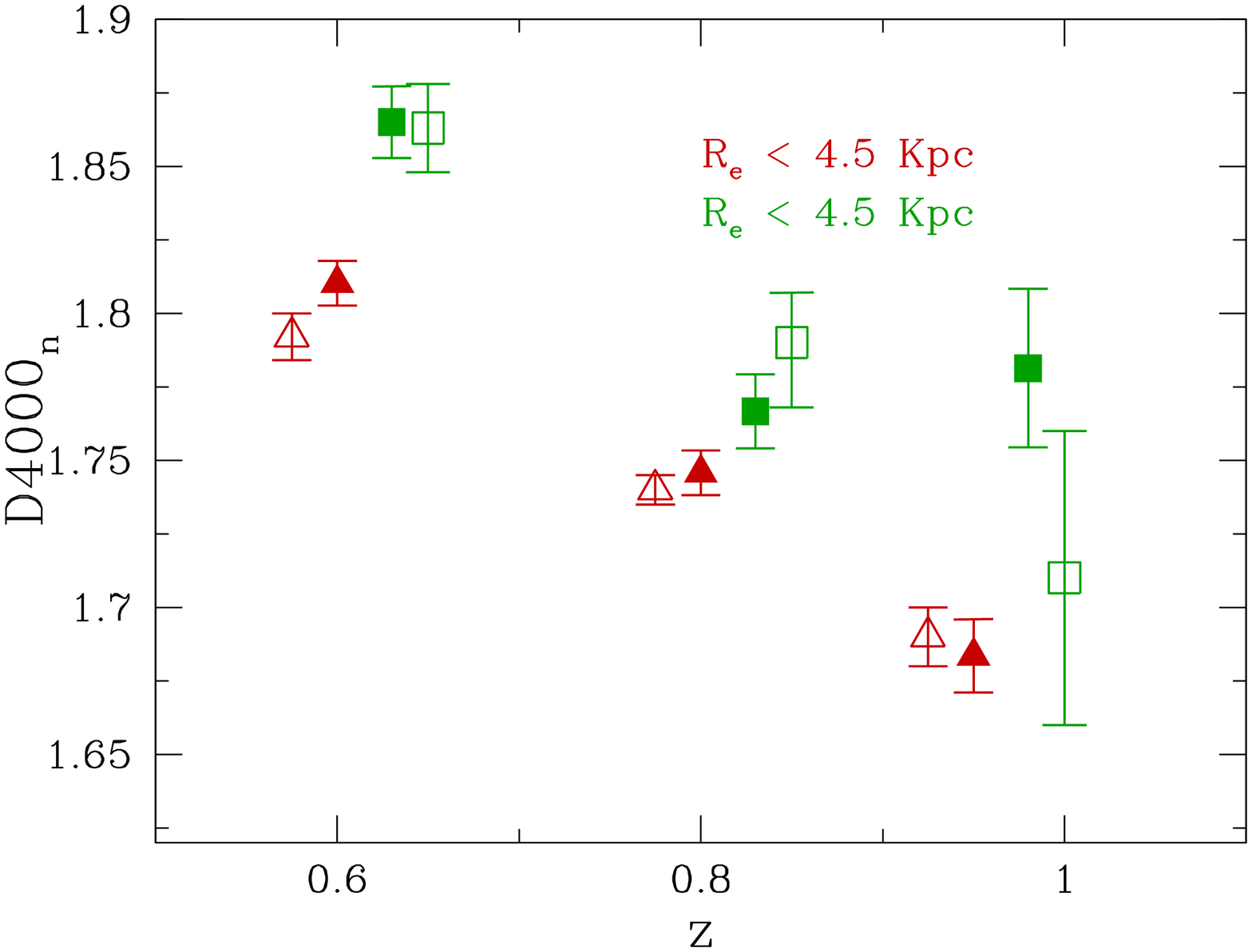}\\
\includegraphics[angle=0,width=4.1cm]{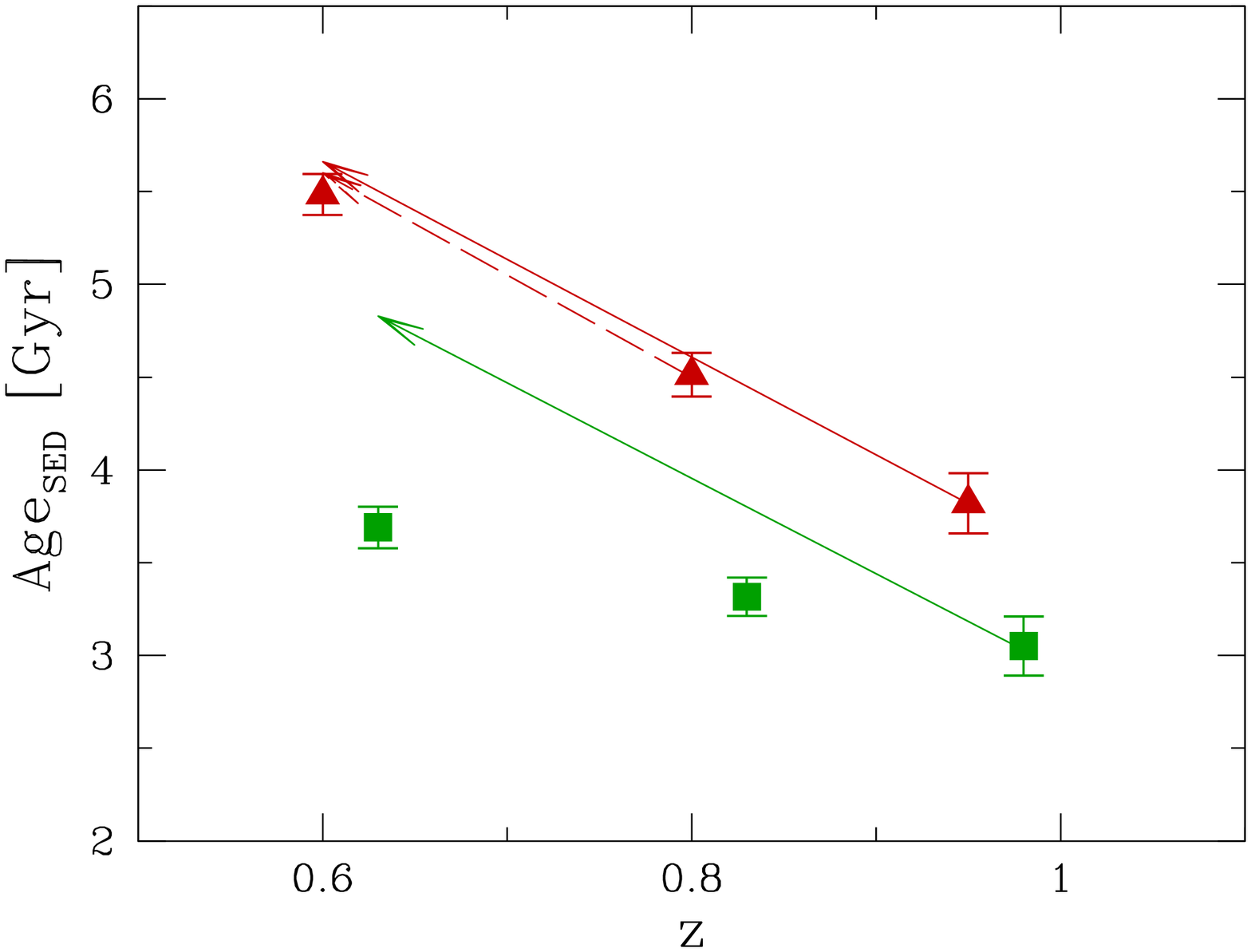} &  
\includegraphics[angle=0,width=4.1cm]{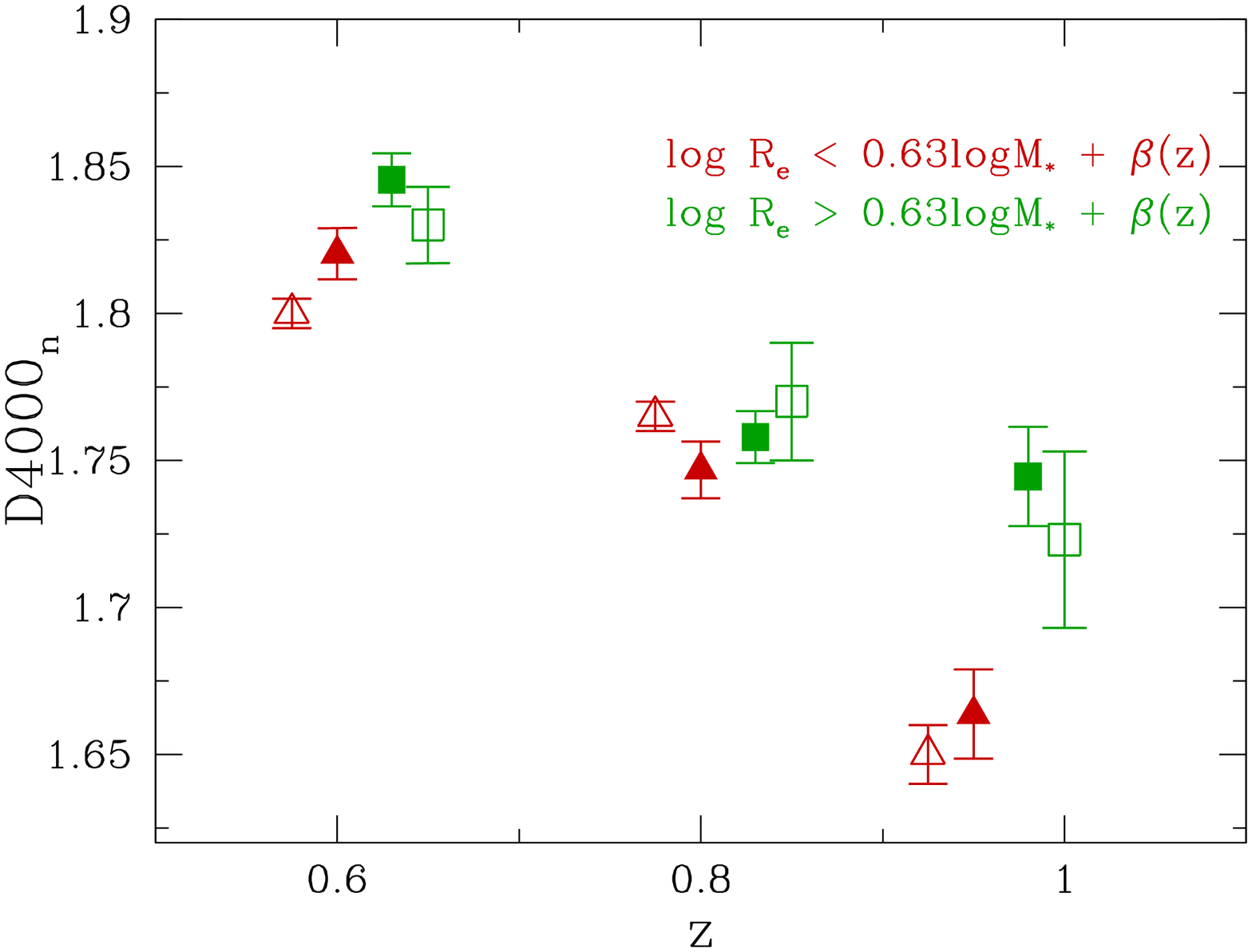}\\
  \end{tabular}
 	\caption{\textit{Upper panels}: the selection criteria of dense and less dense MPGs in this work and in F16 work reported in the size mass plane, in two redshift bins  (0.5 $\leqslant z <$ 0.7 (left panel) and 0.9 $\leqslant z \leqslant$ 1.0 (right panel)). Dashed cyan lines indicate the cut we apply in the analysis presented in this paper (namely $\Sigma$ = 1000\,M$_{\odot}$pc$^{-2}$ and $\Sigma$ = 2000\,M$_{\odot}$pc$^{-2}$). Black lines indicate the two criteria adopted in F16: dashed horizontal lines are the selection at constant $\it{R_e}$, while solid lines are the selection along the SMR such to select the 35$\%$ of the densest (dark red points) and less dense (green points) galaxies in that redshift bin. This selection is redshift dependent. \textit{Middle panels}: the evolution of the age as derived by the fit of the SED (left panel) for MPGs with $\it{R_e} <$ 4.5\,kpc (dark red points) and with $\it{R_e} >$ 7.5\,kpc (green points) (left panel) and the corresponding evolution of the mean D4000$_{n}$ (right panel). In the left panels the solid(dashed) arrow indicate the increase of the age in case of passive evolution since $z = 0.95$(0.8). In the right panel, open symbols indicate the value of D4000$_{n,SED}$ expected from the best-fit values of SED. In both panels error bars indicate the errors on the mean. In the highest redshift bin the largest error for less dense galaxies is principally due to the smaller statistics of the sample ($<$ 30 galaxies) with respect to the others bins ($>$ 100 galaxies). \textit{Lower panels}: the same as middle panels but for MPGs selected along the SMR (red and green points in the upper panels).}
	\label{fag}
 \end{center}
\end{figure}

In this section we repeat the analysis shown in Sect. 5 assuming the same definition for dense and less dense passive galaxies adopted by \citet{fagioli16}. 
As stated in Sect. 5, Fagioli et al. adopted two criteria to select dense/non dense galaxies:
i) cuts at constant $\it{R_e}$ (i.e. $\it{R_e} <$ 4.5\,kpc for dense galaxies and $\it{R_e}>$ 7.5\,kpc for less dense galaxies), and ii) cuts parallel to the SMR. In this second case, the authors fixed the slope to 0.63 and, in any redshift bin, varied the zero-point such to split the sample 
in three sub-populations which account for the 35 : 30 : 35 $\%$ of the total number of galaxies in the redshift bin under consideration. In the upper panels of Fig. \ref{fag}
the selection criteria are shown for the two extreme redshift bins (0.5 $\leqslant z <$ 0.7 in the left panel and 0.9 $\leqslant z \leqslant$ 1.0 in the right panel).

In the middle left panel of Fig. \ref{fag} we plot the evolution of age$_{SED}$ 
for MPGs with R$_{e} <$ 4.5kpc and with R$_{e} >$ 7.5kpc (red and green filled points, respectively). The solid arrows track the evolution for the pure aging of the stellar populations from z = 0.95 to z = 0.6. The figure shows that the age$_{SED}$ of MPGs with  R$_{e} <$ 4.5kpc are consistent with a passive evolution, and that at any z, less dense galaxies are younger. The middle right panel shows the evolution of the mean D4000$_{n}$ for the two sub-populations. 
Open symbols in the right panel indicate the D4000$_{n}$ values expected from the results of the SED fitting (D4000$_{n, SED}$, see Sect. 5 for more details on the procedure). 
The agreement between
D4000$_{n}$ and D4000$_{n, SED}$ is overall good and confirms the results we found in sect. 5 for our sample of dense/less dense MPGs.

Lower panels in Fig. \ref{fag} show the same of the upper panels but for galaxies selected with cut parallel to the SMR. 
Even in this case, the general conclusions do not change.

\section{On the impact of the finite dimension of the slit on the estimate of D4000$_{n}$ for low-$\Sigma$ MPGs}

In Sect. 5 we mentioned  that the 1'' slit of VIPERS spectra just samples the central (r $<$ 0.5$\it{R_e}$) regions 
of low-$\Sigma$ MPGs and up to 1.5$\it{R_e}$ for high-$\Sigma$ MPGs. 
Given the presence of metallicity gradients in passive galaxies, this peculiarity can invalidate
the direct comparison of D4000$_{n}$ for MPGs with different $\Sigma$ and also the comparison between
the stellar ages derived from the fit of the SED (considering the whole galaxy) and 
 the D4000$_{n}$ values for low-$\Sigma$ MPGs.
In this section we want to quantify if and how much the stellar population properties
vary with the radius in passive galaxies, in order to interpret our results at the light of these constraints.

\begin{figure*}
 \begin{center}
\includegraphics[angle=0,width=18.5cm]{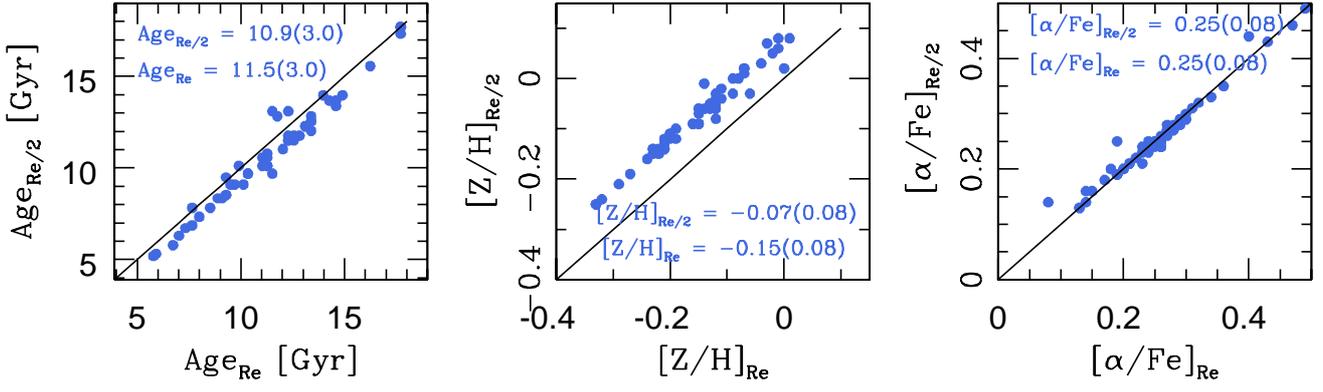}
	\caption{\textit{Left panel}: The stellar population age within 0.5$\it{R_e}$ vs. the stellar population age within
	$\it{R_e}$ for local ETGs with dynamical mass logM$_{JAM}$ $>$ 10.9 and 
	$\Sigma <$ 1000\,M$_{\sun}$pc$^{-2}$ (green points). 
	Solid line is the 1:1 correlation. Mean value of ages at $\it{R_e}$ and at 0.5$\it{R_e}$ are reported 
	in the panel. \textit{Central and right panel}: the same as 
	left panel but for stellar metallicity and alpha enhancement, respectively. }
	\label{atlas}
 \end{center}
\end{figure*}

We address this topic using the spatially resolved information on stellar population properties
provided by ATLAS$^{3D}$ survey \citep{cappellari11} for a sample of local ETGs.
In particular we refer to the estimates of age, metallicity and alpha enhancement [$\alpha$/Fe] within
$\it{R_e}$ and whitin 0.5$\it{R_e}$ presented in \citet{mcdermid15}.
Stellar masses were not provided by the ATLAS$^{3D}$ team; for this reason 
we refer to the dynamical masses M$_{JAM}$ derived as specified in 
Table 1 in \citet{cappellarixv} (M/L$_{JAM}$ $\times$ L, the median dark matter fraction is just 13$\%$, see \citet{cappellarixv}). 
From the same table we took the $\it{R_e}$ values
and derived the mean stellar mass density as M$_{JAM}$/(2$\pi$$\it{R_e}^{2}$). 
From the whole sample of 260 ETGs we selected the massive (log M$_{JAM} > $ 10.9) and less dense galaxies $\Sigma$ ($<$ 1000\,M$_{\sun}$pc$^{-2}$).
We relaxed the mass cut somewhat with respect to the one used in the paper to have a higher statistic (considering that here we are using total mass). In Fig. \ref{atlas} we report the age, metallicity and [$\alpha$/Fe] measured within 0.5$\it{R_e}$ versus their estimates within $\it{R_e}$ for local massive less dense  ETGs.
Figure \ref{atlas} shows that the central regions of local ETGs have approximately the same age and [$\alpha$/Fe] 
of the external regions (Age$_{R_e}$/Age$_{0.5R_e}$ $\simeq$ 1.05, [$\alpha$/Fe$_{R_e}$]/[$\alpha$/Fe$_{0.5R_e}$] $\simeq$ 1.0) 
but are more metal rich (Z$_{0.5R_e}$/Z$_{R_e}$ $\simeq$ 1.2, for similar results see also \citet{spolaor10}). 

\end{appendix} 

\end{document}